\documentclass[12pt]{article}
\usepackage[hmargin=1.5cm, vmargin=2cm]{geometry}
\expandafter\let\csname equation*\endcsname\relax 
\expandafter\let\csname endequation*\endcsname\relax 
\usepackage{amsmath}
\usepackage{amsfonts}
\usepackage{amssymb}
\usepackage{graphicx}
\usepackage{undertilde}
\usepackage{mathrsfs}
\usepackage{bbding}
\usepackage{setspace}
\usepackage{tabularx,ragged2e,booktabs,caption}
\usepackage{booktabs}
\usepackage{makecell}
\usepackage{lscape}
\usepackage{subfigure}
\usepackage{float}
\usepackage{xspace}
\usepackage{url}

{\noindent\ignorespaces}%
{\par\noindent%
\ignorespacesafterend}

\newcommand{\td}{\ensuremath{  (A \rightarrow M)\text{ }   }}
\newcommand{\tdd}{\ensuremath{  (M \rightarrow A)\text{ }   }}
\newcommand{\sxxexx}{{ $\sigma_{11}$--$\epsilon_{11}$ }}
\newcommand{\phit}{{ $\phi$--$t$ }}
\newcommand{\thetat}{{ $\theta$--$t$ }}

\newcolumntype{C}[1]{>{\Centering}m{#1}}

\newcommand{\itemyspace}{\itemsep-3pt}
\begin{document}

\begin{center}
\Large{Three Dimensional Non-Isothermal Ginzburg-Landau \\Phase-Field Model for Shape Memory Alloys}
\end{center}

\normalsize

\begin{center}
R. Dhote$^{1,3}$, M. Fabrizio$^2$, R. Melnik$^3$, J. Zu$^1$

$^1$Mechanical and Industrial Engineering, University 
of Toronto, \\5 King's College Road, Toronto, ON, M5S3G8, Canada\\
$^2$Department of Mathematics, University of Bologna, \\Piazza di 
Porta S. Donato 5, I-40126 Bologna, Italy\\
$^3$M$^2$NeT Laboratory, Wilfrid Laurier University, Waterloo, ON,  
N2L3C5, Canada
\end{center}

\begin{abstract}
In this paper, a macroscopic three dimensional non-isothermal model is proposed to describe hysteresis phenomena and phase transformations in shape memory alloys (SMAs). The model is of phase-field type and is based on the Ginzburg-Landau theory. The hysteresis and phase transformations are governed by the kinetic phase evolution equation using the scalar order parameter, conservation laws of momentum and energy, and a non-linear coupling between stress, strain, and the order parameter in a differential form. One of the important features of the model is that the phase transformation is governed by the stress tensor as opposed to the transformation strain tensor typically used in the literature. The model takes into account different properties of austenite and martensite phases based on the compliance tensor as a function of the order parameter and stress. Representative numerical simulations on a SMA specimen reproduce  hysteretic behaviors observed experimentally in the literature.

\end{abstract}

\section{Introduction}

Over the last few decades, shape memory alloys (SMAs) have attracted increasing attention of physicists, engineers, and applied mathematicians because of their complex microstructures and interesting thermo-mechanical hysteretic behaviors. The SMAs exhibit two unique hysteresis behaviors namely, shape memory effect and pseudoelasticity at lower and higher temperatures (with respect to the threshold temperature), respectively. These behaviors are caused by the underlying atomic rearrangements from a symmetric configuration  (called the austenite ($A$) phase) to another lower symmetric configurations (called the martensite ($M$) phases). Under mechanical and thermal loadings, the atomic rearrangement results in a  macroscopic deformation of a SMA specimen due to diffusionless transformations. The simultaneous occurrence of high stress and high strain properties of SMAs makes them a suitable candidate for actuators and sensors in a wide range of products in automotive, aerospace, medicine and bioengineering applications \cite{Otsuka,Kohl2004,Lagoudas,Miyazaki2009,Ozbulut2011,Elahinia2011}. \medskip 

Several modeling approaches have been proposed to describe hysteretic behaviors in SMAs. A comprehensive overview of different SMA models can be found in, e.g. \cite{Lagoudas, Khandelwal2009, Lagoudas2006}. The approaches based on phenomenology, phase diagram, micromechanics, crystal plasticity, phase-field models, etc. have been described in detail in \cite{Birman1997, Smith2005, Paiva2006,mamivand2013review} and the references therein. In this paper, we focus on the phase-field (PF) model approach. This approach provides a unified framework to describe temperature- and stress- induced transformations.  Several different PF models have been proposed in \cite{Falk1980,Khachaturian1983,Melnik1999,Melnik2000,Artemev2001,Chen2002,Levitas2002a,Levitas2002b,Levitas2003,Ahluwalia2006,Mahapatra2006,Mahapatra2007a,wang2007finite,Bouville2008,Bouville2009,Grandi2012,Dhote2012,Dhote2013}. They differ in free energy description, selection of order parameters (OPs), model formulation and numerical approaches.\medskip 

Inspired by the phase-field modeling of ferroelectric materials, Falk \cite{Falk1980} applied the Landau-Devonshire theory to describe martensitic transformations (MTs) in SMAs by defining the shear strain as an OP. Later, Wang and Khachaturyan \cite{wang1997three} proposed a three dimensional (3D) continuum stochastic-field kinetic model by defining the transformations induced elastic strain as OPs to predict the MTs. Curnoe and Jacobs \cite{curnoe2000twin}, Lookman et al. \cite{Lookman2003}, Bouville and Ahluwalia \cite{Bouville2008,Bouville2009} used an approach by defining elastic strain components as the OPs. The polynomial based phenomenological description of free energy as a function of OPs and their gradients are used to describe the dynamics of phase transformations. One of the previous notable contributions to the PF theory is the Landau free energy proposed by Levitas et al. \cite{Levitas2002a, Levitas2002b, Levitas2003}. The free energy describes the thermo-mechanical properties of different phases using the tensorial OPs. The strain tensor is decomposed into the elastic and the transformational components, where the latter is a function of OPs. The model employs the number of phase evolution equations equal to the number of martenstic variants considered during the phase transformation. Later, Mahapatra and Melnik \cite{Mahapatra2006,Mahapatra2007a} derived a non-isothermal model based on the free energy developed by Levitas et al. \cite{Levitas2002a, Levitas2002b, Levitas2003} by modifying the multi-variant framework to obtain strongly coupled thermo-mechanical models with essential properties of frame indifference and material symmetry. \\

Recently, Berti et al. \cite{Berti2010,Berti2010a}, Grandi et al. \cite{grandi2012macroscale}, Maraldi et al. \cite{maraldi2012non}, and Dhote et al. \cite{Dhote2013} developed a non-isothermal thermodynamic framework to model MTs in SMAs. The macroscopic framework for 1D and 3D models have been developed based on a simplified version of the free energy proposed by Levitas et al. \cite{Levitas2002a}. Here, we are particularly interested in the 3D model proposed in Dhote et al. \cite{Dhote2013}. One of the important features of the 3D model within the non-isothermal framework is the use of a scalar phase OP instead of the tensorial OPs used earlier in the literature \cite{Levitas2002a,Levitas2002b,Levitas2003}. The application of this approach reduces the problem size by limiting the number of phase evolution equations to one, instead of considering separate equations for each crystallographic variant \cite{Levitas2002a,Levitas2002b,Levitas2003}. This results in a simple model which is amenable to an efficient numerical implementation. The other highlights of the model are (i) the rate dependent constitutive equations coupling the stress, strain and phase order parameter, (ii) the description of the phase transformation based on the stress tensor, and (iii) the phase dependent properties by incorporating the compliance tensor based on the local phase value and stress. \medskip 


In this paper, the 3D non-isothermal model is implemented and examples in a 2D setting to study the SMA behavior are provided. In our earlier publication \cite{Dhote2013}, the numerical experiments were conducted on the model by numerically solving the kinetic phase evolution and constitutive equation in 1D and pseudo 2D case driven by stress loading, but without incorporating the conservation laws of momentum and energy. The model is now studied here by incorporating the full thermo-mechanical coupling and phase evolution equations. In addition, the model is simulated with the material properties of the Ni$_ {55} $Ti$_ {45} $ specimen \cite{zhang2010experimental,Grandi2012}. \medskip 

The paper is organized as follows. In Section \ref{sec:generalizedThDmodel}, a 3D non-isothermal model is described using the kinetic phase evolution,  non-linear couplings between stress, strain and order parameter and conservation laws of momentum and energy. In Section \ref{sec:NumericalExpts}, representative simulations on a rectangular SMA specimen are described in detail and studied by using the stress controlled loading. Finally, the conclusions are given in Section \ref{sec:Conclusions}.

\section{3D non-isothermal phase-field model} \label{sec:generalizedThDmodel}

The fully coupled thermo-mechanical PF model is developed to describe the non-linear hysteretic response of SMAs. We define the OP $ \phi $ to describe the austenite ($ \phi = 0 $) and martensite ($ \phi = 1 $) phase. Here we do not distinguish between different variants of martensites. This is a different approach compared to a multi-variant OPs approach (e.g. see \cite{Levitas2002a,Levitas2002b,Levitas2003}). It facilitates the development of a simpler model which is computationally tractable, but at the expense of distinction between different martensitic variants individually. In the following section, the governing equations of the phase evolution and the conservation laws of momentum and energy are described.

\subsection{Phase evolution equation}

\begin{figure}[ht]
\centering
\subfigure[Free energy potentials $\mathscr{F} (\phi)$ and $\mathscr{G} (\phi)$]  
{  
\includegraphics[width=0.35\textwidth]{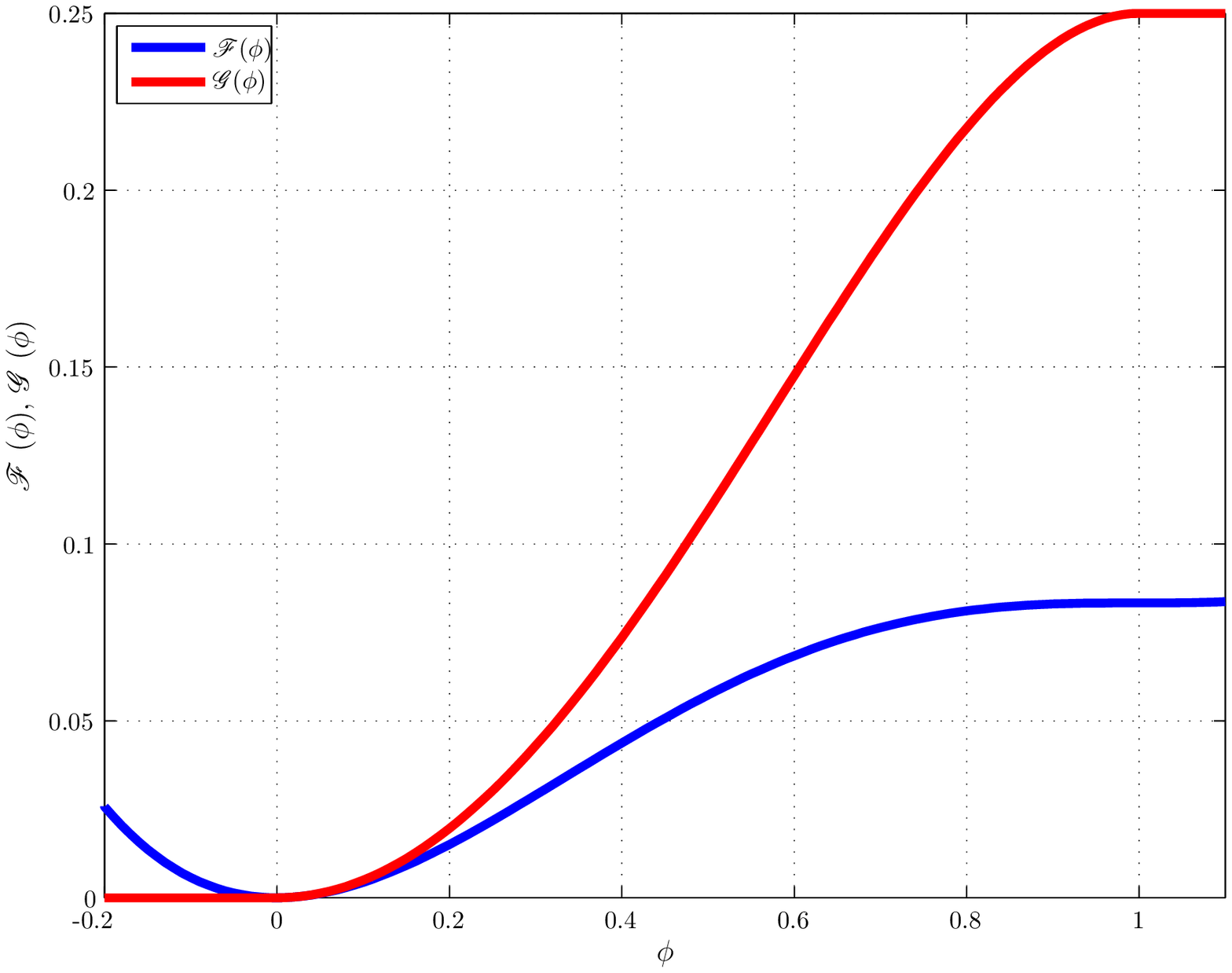}  
\label{fig:figure6a}  
}
\subfigure[Phase space diagram $|\pmb{\sigma}|  \epsilon_0-\theta$ ]  
{ 
\includegraphics[width=0.35\textwidth]{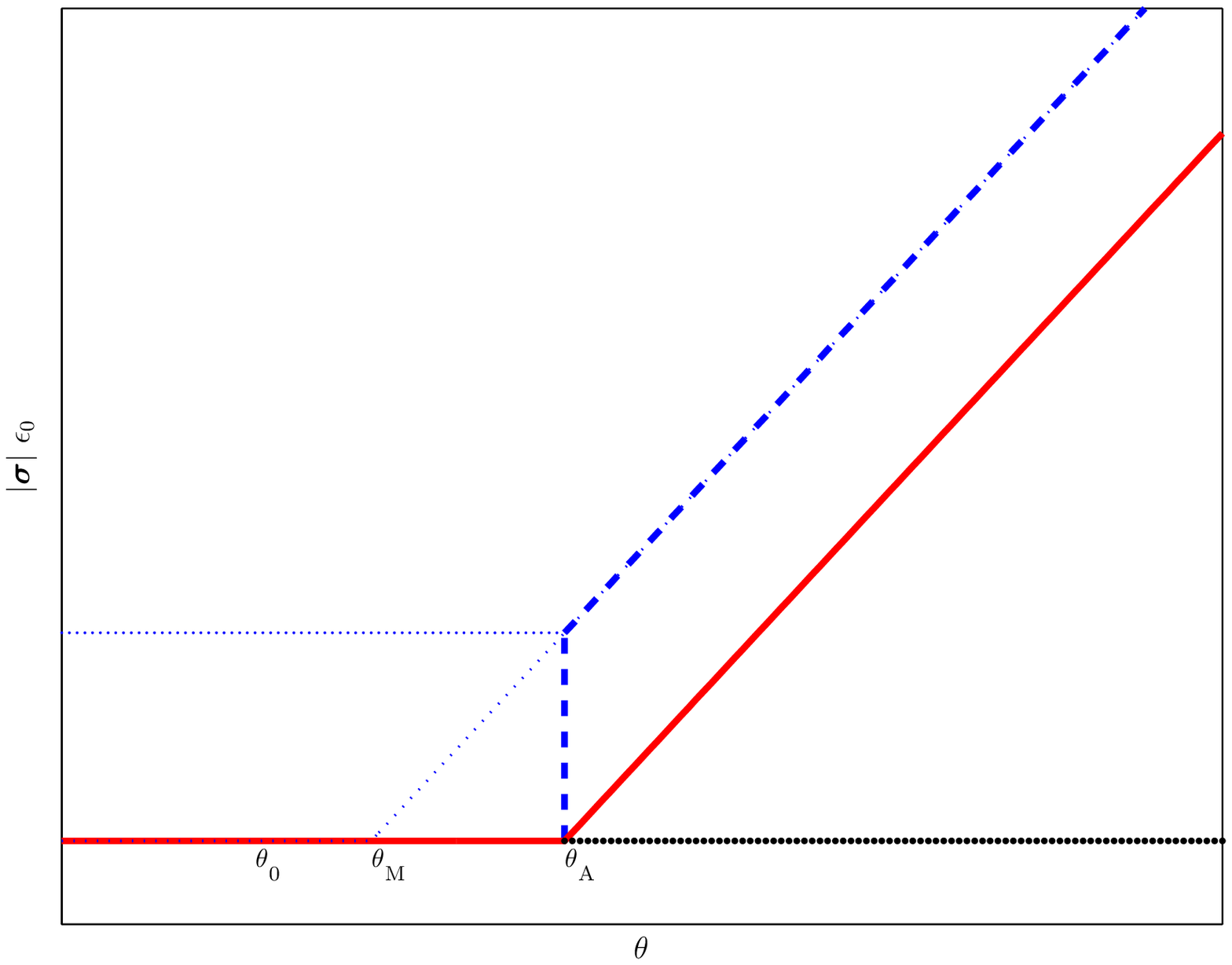}  
\label{fig:figure6b}  
} 
\label{fig:FEUST}
\caption{Free energy $ \mathscr{E} $ and phase space  $|\protect\pmb{\sigma}| \protect\epsilon_0-\protect\theta$ plot.}
\end{figure}

In order to derive a phase evolution equation, we choose a free energy functional having minima at $ \phi = 0 $ and $ \phi = 1 $ with no distinction between martensitic variants. The free energy functional $ \Psi $  based on the Ginzburg-Landau potential is given by
\begin{eqnarray}
\Psi = \frac{\kappa}{2} |\nabla \phi|^2 - \frac{1}{2} (\pmb{\sigma \lambda} 
\cdot \pmb{\sigma}) + \frac{\ell}{2} \left\{  \theta_0 \mathscr{F}(\phi) + \left(\hat{\theta} - \frac{{\epsilon}_0}{\ell} \frac{\pmb{\sigma} \cdot \pmb{\sigma}}{|\pmb{\sigma}|} \right) \mathscr{G}(\phi) \right\},
\label{eq:FEGL}
\end{eqnarray}

where $ \kappa $ is the Ginzburg constant, $ \pmb{\sigma} $ is the stress tensor, $ \pmb{\lambda} $ is the compliance tensor, $ \ell $ is the latent heat of phase transition, $ \theta_0 $ and $ \hat{\theta} $ are the temperatures, $ \epsilon_0 $ is the equivalent transformation strain. The potentials $\mathscr{F}(\phi )$ and $\mathscr{G}(\phi )$ are the 2-3-4  polynomial functions of $ \phi $ defined as
\begin{equation}
\mathscr{F}(\phi )=\frac{1}{2}\phi^{2} -\frac{2}{3}\phi ^{3}+\frac{1}{4}\phi ^{4} +\beta (\phi ^{2}-\phi ),\qquad \mathscr{G}(\phi )=\left\{
\begin{array}{ll}
\displaystyle 0 & \text{if }\phi <0, \vspace{2mm} \\
\displaystyle \frac{1}{2}\phi ^{2}-\frac{1}{4}\phi ^{4}\qquad & \text{if }0\leq \phi \leq 1,  \vspace{2mm}\\
\displaystyle \frac{1}{4} & \text{if }\phi >1.%
\end{array}%
\right.  \label{eq:PotentialsFandG}
\end{equation}%
Here, the constant $\beta $ is a very small perturbing term added to accommodate slope variations in the regime of instability ($0<\beta \ll 1$) as described in \cite{Berti2010a}. We define
\begin{equation}
\hat{\theta}=\left\{
\begin{array}{c l}
\theta -\theta _{A} & \text{if }\theta >\theta _{A}, \\
0 & \text{if }\theta \leq \theta _{A},%
\end{array}%
\right.  \label{eq:switchingcondition}
\end{equation}
where $\theta _{A}>\theta_{0} $ . The temperature $ \theta_M $ is defined as $\theta_M = \theta_A - \theta_0 $. The free energy potentials $  \mathscr{F} $($\phi$) and $ \mathscr{G} $($\phi$) as well as the phase space diagram $|\pmb{\sigma}| \epsilon_0-\theta$, are plotted in Fig. \ref{fig:figure6a} and \ref{fig:figure6b}, respectively. \medskip

The two-well free energy functional $ \mathscr{E} $, defining the minimum at $ \phi = 0 $ and $ \phi=1 $, is mapped to the potentials $ \mathscr{F} $ and $ \mathscr{G} $ as
\begin{equation}
\mathscr{E} =  \mathscr{F} + w \hspace{2mm} \mathscr{G}, 
\end{equation}
where $ w $ is the function of temperature and stress defined as
\begin{equation}
w = \frac{1}{\theta_0}\left(\hat{\theta} - \frac{\epsilon_0}{\ell} \frac{\pmb{\sigma} \cdot \pmb{\sigma}}{|\pmb{\sigma}|}\right).
\end{equation}

\begin{figure}[ht]
\centering
\subfigure[Two-well free energy functional $ \mathscr{E} $]  
{  
\includegraphics[width=0.35\textwidth]{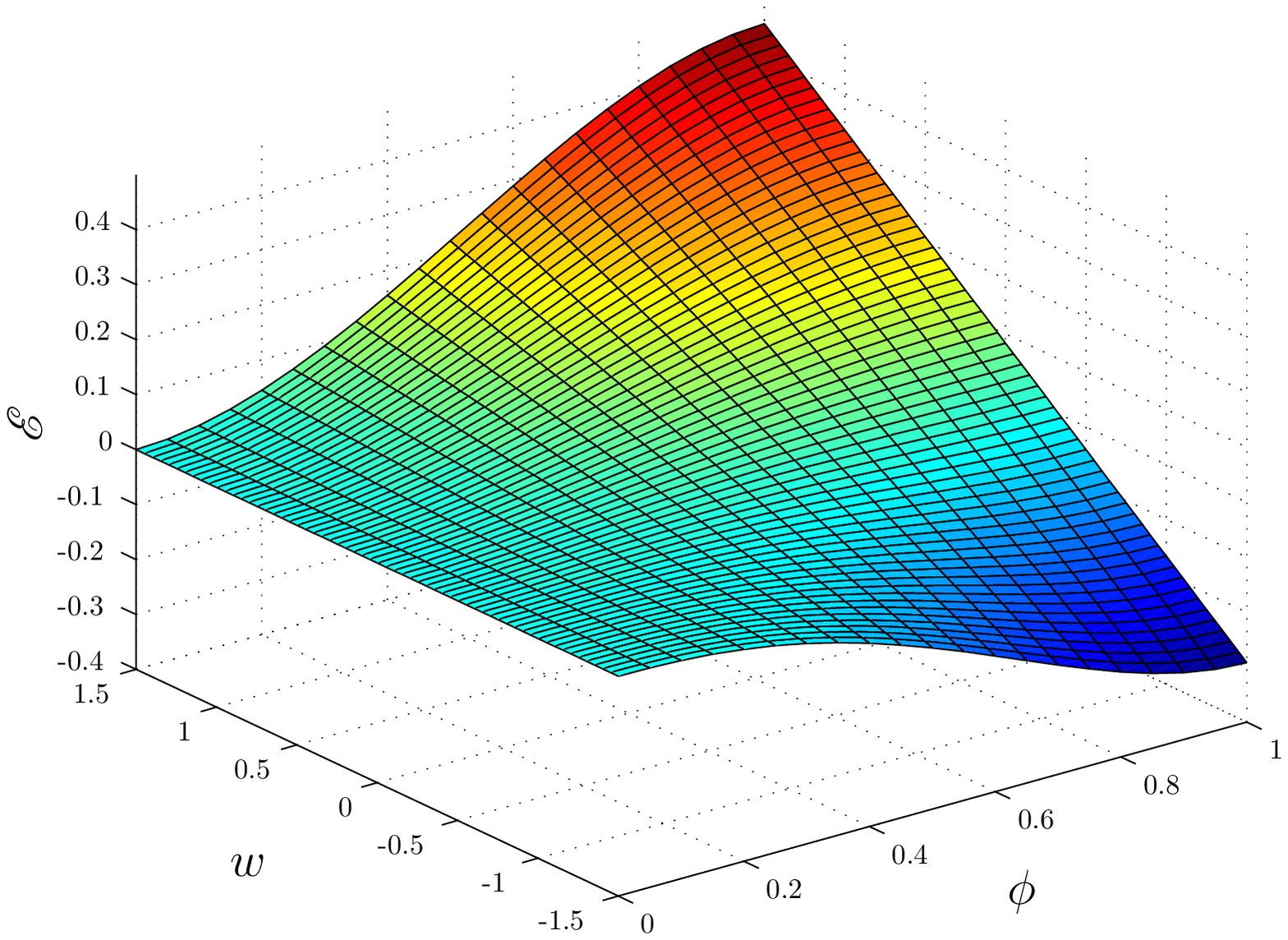}  
\label{fig:TwoWell}  
}
\subfigure[$ \mathscr{E} $ vs $ \phi $ at different values of $ w $]  
{ 
\includegraphics[width=0.35\textwidth]{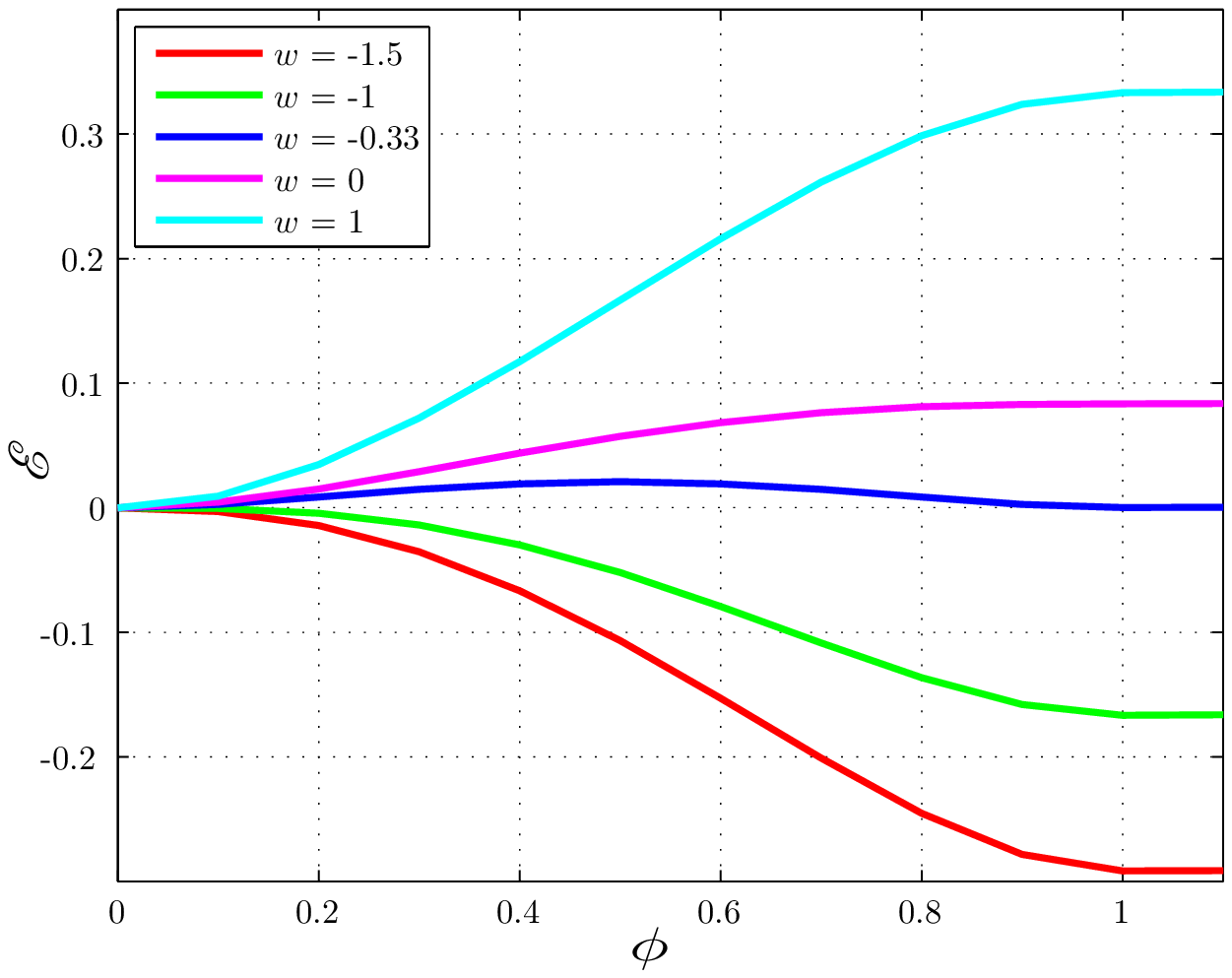}  
\label{fig:Wvsu}  
} 
\caption{Two-well free energy function $ \mathscr{E} $.}
\label{fig:TwoWellFunctional}
\end{figure}

The two-well free energy functional $ \mathscr{E} $ is plotted for different values of $ w $ in Fig. \ref{fig:TwoWellFunctional}. The phase in the domain adheres to the following rules:

\begin{itemize}\itemyspace
\item $ w > 0 $: functional $ \mathscr{E} $ has minimum at $ \phi = 0 $,
\item $ w < -1 $: functional $ \mathscr{E} $ has minimum at $ \phi = 1 $,
\item $ -1 < w < 0 $: functional $ \mathscr{E} $ has metastable states at $ \phi = 0 $ and/or $ \phi = 1 $.
\end{itemize}
Thus the lines $ w = -1 $ and $ w = 0 $ represent the critical threshold for the disappearance of minimum at $ \phi = 0 $ and $ \phi = 1 $, respectively \cite{Berti2010a}. 

We closely follow \cite{Berti2010a,Dhote2013} for deriving the governing equations. For consistency and completeness, the highlights of the main derivation are summarized as follows. 

The temporal evolution of OP is described by the first-order kinetic time-dependent Ginzburg-Landau (TDGL) equation. The TDGL equation is stated as follows:
\begin{equation}
\gamma \frac{\partial \phi}{\partial t} = - \frac{\delta \Psi}{\delta \phi} + \nabla \cdot  \left(\frac{\delta \Psi}{\delta \nabla \phi}\right),
\label{eq:TDGLdef}
\end{equation}
where $ \gamma $ is the relaxation parameter, and $ \delta $ defines the functional derivative. 

On substituting Eq. (\ref{eq:FEGL}) in Eq. (\ref{eq:TDGLdef}) and mathematical manipulation, we obtain the phase evolution equation as
\begin{equation}
\gamma \frac{\partial \phi }{\partial t}=\kappa \Delta \phi - \frac{\ell}{2} \left\{ \theta _{0} \frac{\partial \mathscr{F}(\phi )}{\partial \phi } + \left( \hat{\theta}-\frac{{\epsilon}_{0}}{\ell}\xspace\frac{\pmb{\sigma}\xspace\cdot \pmb{\sigma}\xspace}{|\pmb{\sigma}\xspace|}\right) \frac{\partial \mathscr{G}(\phi )}{\partial \phi } \right\}.
\label{eq:strainrate}
\end{equation}%

\subsection{Structural equation}

The structural equations are described by using the kinematic relationship, appropriate constitutive equation and the conservation law of momentum. 

\subsubsection{Kinematic relationship}
The model is developed based on isotropic material properties and small strain framework. The infinitesimal Cauchy-Lagrange strain tensor $\pmb{\epsilon}$ is defined as
\begin{equation}
\pmb{\epsilon} = \frac{1}{2} \left( \nabla \pmb{u} + \nabla \pmb{u}^T  \right),
\label{eq:Kinematicrel}
\end{equation}
where $\pmb{u}$ is the displacement vector, and $ \pmb{x} $ is the spatial coordinate vector.

\subsubsection{Constitutive relationship}

The relationship between the stress and strain is defined using the constitutive equations. In the case of the phase changing systems, the constitutive equations are modified to account for the phase transformation to describe hysteresis. Thus the constitutive equations not only relate the $\pmb{\sigma}$ and $\pmb{\epsilon}$, but also the $\phi$. The common methodology to achieve this is to decompose strain into the elastic and phase transformation components \cite{Khachaturian1983,Levitas2002a,Levitas2002b,Levitas2003,Mahapatra2006,Grandi2012,maraldi2012non}. We too follow this methodology, however, we allow to account the different forms of dependencies of compliance on $\phi$ and $\pmb{\sigma}$. In this case, the constitutive equation is of the differential form \cite{Berti2010a,Dhote2013} described as

\begin{equation}
\dot{ \pmb{\epsilon}_{} \xspace} = \pmb{\lambda}_{} \xspace^{1/2}(
\pmb{\sigma} \xspace,\phi) \frac{\partial}{\partial t} \left( \pmb{\lambda}
_{} \xspace^{1/2}(\pmb{\sigma} \xspace,\phi) \pmb{\sigma} \xspace \right) + 
{\epsilon}_{0} \xspace \frac{\pmb{\sigma} \xspace}{|\pmb{\sigma} \xspace|
} \dot{G}(\phi).  \label{eq:OPpde3d}
\end{equation}

The first and second terms in Eq. (\ref{eq:OPpde3d}) represent the elastic and transformational components of the strain in a differential form. The compliance tensor $\pmb{\lambda}_{} \xspace$ accommodates the properties of both austenite and martensite phases based on the OP value and stress \cite{Levitas2002a}. In the general $3$D case, it is defined as

\begin{equation}
\pmb{\lambda}_{} \xspace(\pmb{\sigma} \xspace,\phi) = \pmb{\lambda}_{2} 
\xspace(\phi) + \pmb{\lambda}_{3} \xspace(\phi) \pmb{\sigma} \xspace + 
\pmb{\lambda}_{4} \xspace(\phi) \pmb{\sigma} \xspace \cdot \pmb{\sigma} 
\xspace,  \label{eq:lambda}
\end{equation}

where $\pmb{\lambda}_{2}$\xspace, $\pmb{\lambda}_{3}$\xspace, 
$\pmb{\lambda}_{4}$\xspace are the tensor of $4^{th}$, $6^{th}$, and $8^{th}$ order, respectively. A point to be noted in Eq. (\ref{eq:OPpde3d}) that the phase transformation is governed by the stress tensor $ \pmb{\sigma} $ as opposed to the transformation strain tensor described, e.g., in  \cite{Khachaturian1983,Levitas2002a,Levitas2002b,Levitas2003,Mahapatra2006,Grandi2012,maraldi2012non}.

\subsubsection{Conservation of momentum}

The structural balance is governed by the conservation law of momentum. It reads as
\begin{equation}
\rho \ddot{\mathbf{\pmb{u} }} = \nabla \cdot \pmb{\sigma} \xspace + \rho \mathbf{\pmb{f} },  \label{eq:momentumbalance}
\end{equation}
where $\rho$ is the density, and \textbf{$\pmb{f}$} is the body force. 

\subsection{Thermal equation and thermodynamic consistency}

The free energy $ \psi $ can be expressed in the terms of internal energy $ e $ and entropy $ \eta $ as
\begin{equation}
\psi = e - \theta \eta.
\label{eq:freeenergy}
\end{equation}

According to the first law of thermodynamics, the balance of energy is written as
\begin{equation}
\rho\dot{e}(\pmb{\sigma}\xspace,\phi ,\theta )=\mathscr{P}_{m}^{i}+
\mathscr{P}_{\phi }^{i}- \nabla \cdot \mathbf{\pmb{q} } + r,
\label{eq:firstTD}
\end{equation}%
where $\mathscr{P}_{m}^{i}$ is the internal mechanical power, $\mathscr{P}_{\phi }^{i}$ is the internal order structure power, $\pmb{q}$ is the heat flux defined as $\pmb{q} = - k \nabla \theta$ and $r$ is the external heat source. The $\mathscr{P}_{m}^{i}$ is defined as
\begin{eqnarray}
\mathscr{P}_{m}^{i} &=&\pmb{\sigma}\xspace\cdot \dot{\pmb{\epsilon}_{{}}
\xspace},  
\end{eqnarray}
and the $\mathscr{P}_{\phi }^{i}$ is obtained by multiplying Eq. (\ref{eq:strainrate}) by $\dot{\phi}$ and following the approach of \cite{Berti2010} 
\begin{eqnarray}
\mathscr{P}_{\phi }^{i} &=&\gamma \dot{\phi}^{2}+ \frac{\kappa}{2} 
\frac{d}{dt}\left( |\nabla \phi |^{2}\right) + \frac{\ell}{2} \left\{ \theta _{0}\dot{\mathscr{F}}(\phi )+\left( 
\hat{\theta}-\frac{{\epsilon}_{0}}{\ell}\xspace\frac{\pmb{\sigma}\xspace\cdot 
\pmb{\sigma}\xspace}{|\pmb{\sigma}\xspace|}\right) \dot{\mathscr{G}}(\phi )\right\}.  
\label{eq:IntPowers}
\end{eqnarray}%

In order to prove consistency of the above model to the second law of thermodynamics, the Clausius-Duhem inequality needs to be satisfied. Thus 
\begin{equation}
\dot{\eta} \geq - \nabla \cdot \left( \frac{\pmb{q}}{\theta} \right) + 
\frac{r}{\theta}.
\label{eq:secondlaw}
\end{equation}

Using Eqs. (\ref{eq:freeenergy})-(\ref{eq:IntPowers}) and (\ref{eq:secondlaw}), the inequality is reduced to 

\begin{equation}
\dot{\psi} + \dot{\theta} \eta \le - \frac{\pmb{q}}{\theta} \cdot \nabla 
\theta + \frac{1}{2}\frac{d}{dt}\left( \pmb{\lambda}_{{}}\xspace(\pmb{\sigma} 
\xspace,\phi )\pmb{\sigma}\xspace\cdot \pmb{\sigma}\xspace\right) + \gamma 
\dot{\phi}^{2} + \kappa \nabla \phi  \cdot \nabla \dot{\phi}
+ \frac{\ell}{2} \theta _{0}\dot{\mathscr{F}}(\phi )+ \frac{\ell}{2} \hat{\theta} \dot{\mathscr{G}}(\phi).
\label{eq:secondlawsimplifythree}  
\end{equation}

Under the assumption that the free energy $ \psi $ is a function of the variables $ \phi, \nabla \phi, \pmb{\sigma}, \theta $, Eq. (\ref{eq:secondlawsimplifythree}) can be written as
\begin{eqnarray}
&&\left( \eta + \frac{\partial \psi}{\partial \theta} \right) \dot{\theta} 
+ \left( \frac{\partial \psi}{\partial \phi} - \frac{\ell}{2} \theta _{0}\mathscr{F}^{\prime}(\phi) - \frac{\ell}{2} \hat{\theta} \mathscr{G}^{\prime}(\phi) \right) \dot{\phi}
+ \left( \frac{\partial \psi}{\partial \nabla \phi} - \kappa \nabla \phi 
\right) \cdot \nabla \dot{\phi}  \nonumber \\
&& + \left( \frac{\partial \psi}{\partial \pmb{\sigma}} -  
\pmb{\lambda}_{{}}\xspace(\pmb{\sigma}\xspace,\phi )\pmb{\sigma} \right) \cdot 
\dot{\pmb{\sigma}} - \gamma \dot{\phi}^2 + \frac{\pmb{q}}{\theta} \cdot \nabla 
\theta \le 0.
\label{eq:secondlawsimplified}
\end{eqnarray}

The above inequality is satisfied for the arbitrariness of $ \dot{\phi}, \nabla \dot{\phi}, \pmb{\dot{\sigma}}, \dot{\theta} $ under the following constitutive relations
\begin{eqnarray}
\eta = - \frac{\partial \psi}{\partial \theta}, \qquad
\frac{\partial \psi}{\partial \phi} = \frac{\ell}{2} \left[\theta _{0}\mathscr{F}^{\prime}(\phi) + \hat{\theta} \mathscr{G}^{\prime}(\phi)\right], \qquad 
\frac{\partial \psi}{\partial \nabla \phi} = \kappa \nabla \phi, \qquad
\frac{\partial \psi}{\partial \pmb{\sigma}} =  
\pmb{\lambda}_{{}}\xspace(\pmb{\sigma}\xspace,\phi )\pmb{\sigma},
\label{eq:simplifications}
\end{eqnarray}
where $ \mathscr{F}^{\prime} $ and $ \mathscr{G}^{\prime} $ are derivatives with respect to OP $ \phi $. \newline
Substituting Eq. (\ref{eq:simplifications}) in (\ref{eq:secondlawsimplified}) leads to 
\begin{equation}
\frac{k}{\theta} |\nabla \theta |^2 + \gamma \dot{\phi}^2 \ge 0,
 \end{equation}
thus proving the thermodynamic consistency with the positivity of the thermal conductivity $ k $.

The relations in Eq. (\ref{eq:simplifications}) enforce the following representation of the free energy:
\begin{equation}
\psi = \psi_0(\theta) + \frac{1}{2}\left( 
\pmb{\lambda}_{{}}\xspace(\pmb{\sigma} \xspace,\phi )\pmb{\sigma}\xspace\cdot 
\pmb{\sigma}\xspace\right) + \frac{\kappa}{2}  |\nabla \phi |^{2} + \frac{\ell}{2} \left[\theta_{0} \mathscr{F}(\phi)+ \hat{\theta} \mathscr{G}(\phi)\right],
\label{eq:FreeEnergyEqn}
\end{equation}
where we choose $ \psi_0 $ as
\begin{eqnarray}
\psi_0(\theta) = - \frac{c}{2 \theta_{c}} \theta^2.
\end{eqnarray}

Now, the expression of the internal energy $e$ in Eq. (\ref{eq:freeenergy}) can be simplified as
\begin{eqnarray}
e = \frac{c}{2 \theta_c} \theta^2 + \frac{1}{2} \left( 
\pmb{\lambda}_{{}}\xspace(\pmb{\sigma} \xspace,\phi )\pmb{\sigma}\xspace\cdot 
\pmb{\sigma}\xspace\right) + \frac{\kappa}{2}  |\nabla \phi |^{2} + \frac{\ell}{2}
\left[\theta _{0} \mathscr{F}(\phi) + (\hat{\theta} - \theta \hat{\theta}^{\prime}) \mathscr{G}(\phi)\right],
\label{eq:ieone}
\end{eqnarray}
where $ \hat{\theta}^{\prime} $ is the derivative with respect to temperature $ \theta $. 

On differentiating Eq. (\ref{eq:ieone}) with respect to time and equating it to the right hand side of Eq. (\ref{eq:firstTD}), we obtain the heat equation as
\begin{eqnarray}
\frac{c}{\theta_c} \theta \dot{\theta} - \frac{\ell}{2} \theta \left[ \mathscr{G}(\phi) \hat{\theta}^{\prime \prime} \dot{\theta} + \hat{\theta}^{\prime} \dot{\mathscr{G}}(\phi) \right] - 
\gamma \dot{\phi}^2 = k \Delta \theta + r.
\label{eq:heateqn}
\end{eqnarray}

Defining $\hat{\theta}^{\prime}$ and $\hat{\theta}^{\prime \prime}$ as the Heaviside function $H$ and the Dirac-delta function $\delta_d$, the Eq. (\ref{eq:heateqn}) can be written explicitly as
\begin{eqnarray}
\frac{c}{\theta_c} \theta \dot{\theta} - \frac{\ell}{2} \theta \left[ \mathscr{G}(\phi) \delta_d \dot{\theta} +  H \dot{\mathscr{G}}(\phi) \right] - 
\gamma \dot{\phi}^2 = k \Delta \theta + r.
\label{eq:heateqn1}
\end{eqnarray}
\subsection{System of equations and boundary conditions}

Now, we summarize all the governing partial differential equations and constitutive relations for the above model as

\begin{subequations}
\renewcommand{\theequation}{\theparentequation.\arabic{equation}}
\label{eq:energyfuncationals1}
\begin{eqnarray}
&& \gamma \frac{\partial \phi }{\partial t}=\kappa \Delta \phi - \frac{\ell}{2} \left\{ \theta _{0} \frac{\partial \mathscr{F}(\phi )}{\partial \phi } + \left( \hat{\theta}-\frac{{\epsilon}_{0}}{\ell}\xspace\frac{\pmb{\sigma}\xspace\cdot \pmb{\sigma}\xspace}{|\pmb{\sigma}\xspace|}\right) \frac{\partial \mathscr{G}(\phi )}{\partial \phi } \right\},
\label{eq:gov1} \\
&&\rho \ddot{\mathbf{\pmb{u} }} = \nabla \cdot \pmb{\sigma} \xspace + \rho \mathbf{\pmb{f} },\label{eq:gov2} \\
&&\frac{c}{\theta_c} \theta \dot{\theta} - \frac{\ell}{2} \theta \left[ \mathscr{G}(\phi) \delta_d \dot{\theta} + H \dot{\mathscr{G}}(\phi) \right] - \gamma \dot{\phi}^2 = k \Delta \theta + r. \label{eq:gov3}
\end{eqnarray}
\end{subequations}

\begin{figure}[h!]
\centering
\includegraphics[width=0.4\textwidth]{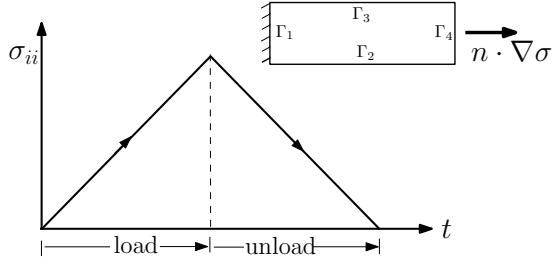}
\caption{Schematic of boundary nomenclature and ramp loading and unloading}
\label{fig:Loading}
\end{figure}

The kinematic relations and constitutive equation are described as
\begin{subequations}
\renewcommand{\theequation}{\theparentequation.\arabic{equation}}
\label{eq:energyfuncationals2}
\begin{eqnarray}
&&\displaystyle\mathbf{\pmb{\epsilon} } = \frac{1}{2} \left[ \nabla \mathbf{\pmb{u} } + \nabla \mathbf{\pmb{u} }^T \right], \\
&&\displaystyle\dot{ \pmb{\epsilon}_{} \xspace} = \pmb{\lambda}_{} \xspace^{1/2}( \pmb{\sigma} \xspace,\phi) \frac{\partial}{\partial t} \left( \pmb{\lambda}_{} \xspace^{1/2}(\pmb{\sigma} \xspace,\phi) \pmb{\sigma} \xspace \right) + \pmb{\epsilon}_{0} \xspace \frac{\pmb{\sigma} \xspace}{|\pmb{\sigma} \xspace|} \dot{\mathscr{G}}(\phi),
\end{eqnarray}
\end{subequations}

along with the boundary conditions (refer to Fig. \ref{fig:Loading}
for the boundary nomenclature)
\begin{eqnarray}
\pmb{n} \cdot \nabla \phi \big|_{\pmb{\Gamma}} = 0, \quad 
\pmb{u}\big|_{\Gamma_1}  = \pmb{0}, \quad 
\pmb{n} \cdot \nabla \pmb{\sigma} \big|_{\Gamma_{2,3}} = 0,\quad 
\pmb{n} \cdot \nabla \pmb{\sigma} \big|_{\Gamma_{4}} = \bar{\pmb{\sigma}},\quad 
\pmb{n} \cdot \nabla \theta\big|_{\pmb{\Gamma}} = h (\theta - \theta_{\text{ext}}),
\end{eqnarray}
where $ h $ is the heat transfer coefficient and $ \theta_{\text{ext}} $ is the external environment temperature. The initial conditions are defined as
\begin{eqnarray}
\phi(\pmb{x},0) = \phi_0(\pmb{x}), \qquad \pmb{u}(\pmb{x},0) = \pmb{u}_0(\pmb{x}), \qquad \dot{\pmb{u}}(\pmb{x},0) = \dot{\pmb{u}}_0(\pmb{x}), \qquad \theta(\pmb{x},0) = \tilde{\theta}_{0}(\pmb{x})= \theta_{\text{ext}}(\pmb{x}).
\end{eqnarray}

\subsection{Weak formulation} \label{sec:WeakForm}
Let $ \Omega \subset  \mathbb{R}^d$ be an open set in the $d$-dimensional space  ($d$ = 2,3) defined by the coordinate system $ \pmb{x} $. The boundary is denoted by $ \pmb{\Gamma} $ and its outward normal by $\boldsymbol{n}$. As the constitutive equations are assumed in the differential form in Eqs. (\ref{eq:energyfuncationals2}), we solve them along with Eqs. (\ref{eq:energyfuncationals1}). The weak formulation of Eqs. (\ref{eq:gov1})-(\ref{eq:gov3}) and (\ref{eq:energyfuncationals2}) are derived by multiplying the equations with weighing functions $ \{ \Phi, \pmb{U}, \pmb{\Sigma}, \Theta \} $ and transforming them by using the integration by parts. Let $ X $ denote both the trial solution and weighting function spaces, which are assumed to be identical. Let $ \left(\cdot,\cdot\right)_{\Omega} $ denote the $L^2$ inner product with respect to the domain $ \Omega $. The variational formulation is stated as follows:\\
Find solution $\boldsymbol{S} = \left\{\phi,\boldsymbol{u}, \pmb{\sigma}, \theta \right\} \in X$ such that 
$ \forall \boldsymbol{W}= \left\{\Phi,\boldsymbol{U}, \pmb{\Sigma}, \Theta 
\right\} \in X:$\\
B($\boldsymbol{W},\boldsymbol{S}$) = 0,
with
\begin{align}
B(\boldsymbol{W},\boldsymbol{S})& &= 
\left( \Phi, \gamma \dot{\phi} \right)_{\Omega} 
+ \left( \nabla \cdot \Phi, \kappa \displaystyle \nabla \cdot \phi \right)_{\Omega} 
+ \left( \Phi,  \frac{\ell}{2} \theta_0 \frac{\partial \mathscr{F}}{\partial \phi}  \right)_{\Omega} 
+ \left( \Phi, \frac{\ell}{2} \hat{\theta} \frac{\partial \mathscr{G}}{\partial \phi} \right)_{\Omega} 
- \left( \Phi, \epsilon_0 \frac{\pmb{\sigma} \cdot \pmb{\sigma}}{|\pmb{\sigma}|} \frac{\partial \mathscr{G}}{\partial \phi} \right)_{\Omega} 
\nonumber \\
&&\displaystyle+ \left( \boldsymbol{U}, \rho \boldsymbol{\ddot{u}} \right)_{\Omega} 
+ \left( \nabla \cdot \boldsymbol{U},  \boldsymbol{\sigma} \right)_{\Omega}
- \left( \boldsymbol{U}, \rho \boldsymbol{f} \right)_{\Omega}  
+ \left( \pmb{\Sigma}, \dot{\pmb{\epsilon}} \right)_{\Omega}
- \left( \pmb{\Sigma}, \pmb{\lambda}_{} \xspace^{1/2} \frac{\partial}{\partial t} \left( \pmb{\lambda}_{} \xspace^{1/2} \pmb{\sigma} \xspace \right) \right)_{\Omega}
\nonumber \\
&& \displaystyle
- \left( \pmb{\Sigma}, \pmb{\epsilon}_{0} \xspace \frac{\pmb{\sigma} \xspace}{|\pmb{\sigma} \xspace|} \dot{\mathscr{G}}(\phi) \right)_{\Omega}
+ \left( \boldsymbol{\Theta} , c \frac{\theta}{\theta_c} \dot{\theta} \right)_{\Omega} 
- \left( \Theta, \frac{\ell}{2} \theta \dot{\theta} \delta_d  \mathscr{G} \right)_{\Omega} 
- \left( \Theta, \frac{\ell}{2} \theta H \dot{\mathscr{G}} \right)_{\Omega} 
- \left( \Theta, \gamma \dot{\phi}^2 \right)_{\Omega} 
\nonumber \\
&& \displaystyle 
+ \left( \nabla \cdot \Theta, k \nabla \cdot \theta \right)_{\Omega}
- \left( \Theta, r \right)_{\Omega} 
-\left( \Phi, \kappa \nabla \phi \cdot \pmb{n} \right)_{\Gamma}
-\left( \pmb{U}, \pmb{\sigma} \cdot \pmb{n} \right)_{\Gamma}
-\left( \Theta, k \nabla \theta \cdot \pmb{n} \right)_{\Gamma}.
\label{eq:WeakForms} 
\end{align}

The Eqs. (\ref{eq:WeakForms}) have been implemented in a weak finite element formulation in the Comsol Multiphysics software \cite{Comsol}. 

\section{Numerical experiments} \label{sec:NumericalExpts}

We exemplify the SMA hysteretic behavior in a 2D setting. A rectangular specimen of domain $\Omega = [0,l_x] \times [0,l_y]$ is chosen for the numerical simulations. We assume the plane stress formulation with a consideration that the thickness is small as compared to the other dimensions. For simplicity, we adopt the quasistatic approximation under the assumption that the timescales of thermal dynamics and phase evolution phenomenon are larger than the stress wave time scale \cite{Grandi2012,maraldi2012non}. We neglect the non-linear effect of $\pmb{\lambda}_3$ and $\pmb{\lambda}_4$ compliance tensors. We assume that no external body and thermal loads are applied during the simulation. The phase-dependent compliance tensor $\pmb{\lambda}_2$ takes the form 
\begin{eqnarray}
\pmb{\lambda}_2 = 
\lambda_{A}
\begin{bmatrix}
\displaystyle \left( 1 + \Xi \phi \right) & \displaystyle - \mu \left( 1 + \Xi \phi \right) & 0 \\ 
\displaystyle - \mu \left( 1 + \Xi \phi \right) & \displaystyle \left( 1 + \Xi \phi \right) & 0 \\ 
0 & 0 & \displaystyle (1+\mu)\left( 1 + \Xi \phi \right)
\end{bmatrix},
\label{eq:planestress}
\end{eqnarray}
where $ \Xi = (\lambda_{M}-\lambda_{A})/\lambda_{A} $. The material properties of Ni$_{55}$Ti$_{45}$ specimen \cite{zhang2010experimental,Grandi2012} are summarized in Table \ref{tab:materialparameters}. The governing equations are first rescaled and then implemented in the weak formulation. The results are later converted back to the dimensional form.  \medskip

\begin{minipage}{\linewidth}
\centering
\captionof{table}{Material parameters of polycrystalline Ni$_{55}$Ti$_{45}$} \label{tab:materialparameters} 
\begin{tabular}{ C{3.5cm} C{3.5cm} C{2.5cm} C{1.5cm} C{1.5cm} }\toprule[1.5pt]
$\lambda_A$ & $\lambda_M$ &   $\ell$  & $ \epsilon_0 $ & $\theta_A$ \\\midrule
$2.5\times10^{-11}$ Pa$^{-1}$ & $3.57\times10^{-11}$ Pa$^{-1}$ & $10^{6}$ Pa-K$^{-1}$ & 0.14  & 288.5 K  \\ 
\bottomrule[1.25pt] 
\end {tabular}\par
\bigskip
\begin{tabular}{ C{1.5cm} C{1.5cm}  C{1.2cm}  C{3.0cm} C{3.0cm} C{1.5cm} }\toprule[1.5pt]
$\theta_M$ & $\theta_0$ & $\theta_{c}$ & $c$ &  $k$ &$\kappa$  \\ \midrule
273 K & 212.7 K & 296  K   & $3.2 \times 10^{6}$ Pa-K$^{-1}$ &  18 Wm$^{-1}$K$^{-1}$ &  0.15 N \\  
\bottomrule[1.25pt] 
\end {tabular}\par
\bigskip
\end{minipage}

%

In the following subsections, we describe the results of  the simulations that have been carried out on a rectangular domain with $l_x$ = 0.1 m and $l_y$ = 0.008 m to show the ability of the model to reproduce the SMA hysteretic behavior under the stress-controlled loading. It should be noted that the stress-controlled loading often leads to uniform nucleation in a specimen, which is different from non-uniform nucleation observed during a displacement-controlled loading \cite{zhang2010experimental,Grandi2012}. Hence, a stress-drop is not observed during the phase nucleation in a stress-controlled loading \cite{maraldi2012non}. 
\subsection{Phase-dependent properties} \label{sec:phasedepprop}

\begin{figure}[H]
\centering
\subfigure[\phit]
{
\includegraphics[width=0.31\linewidth]{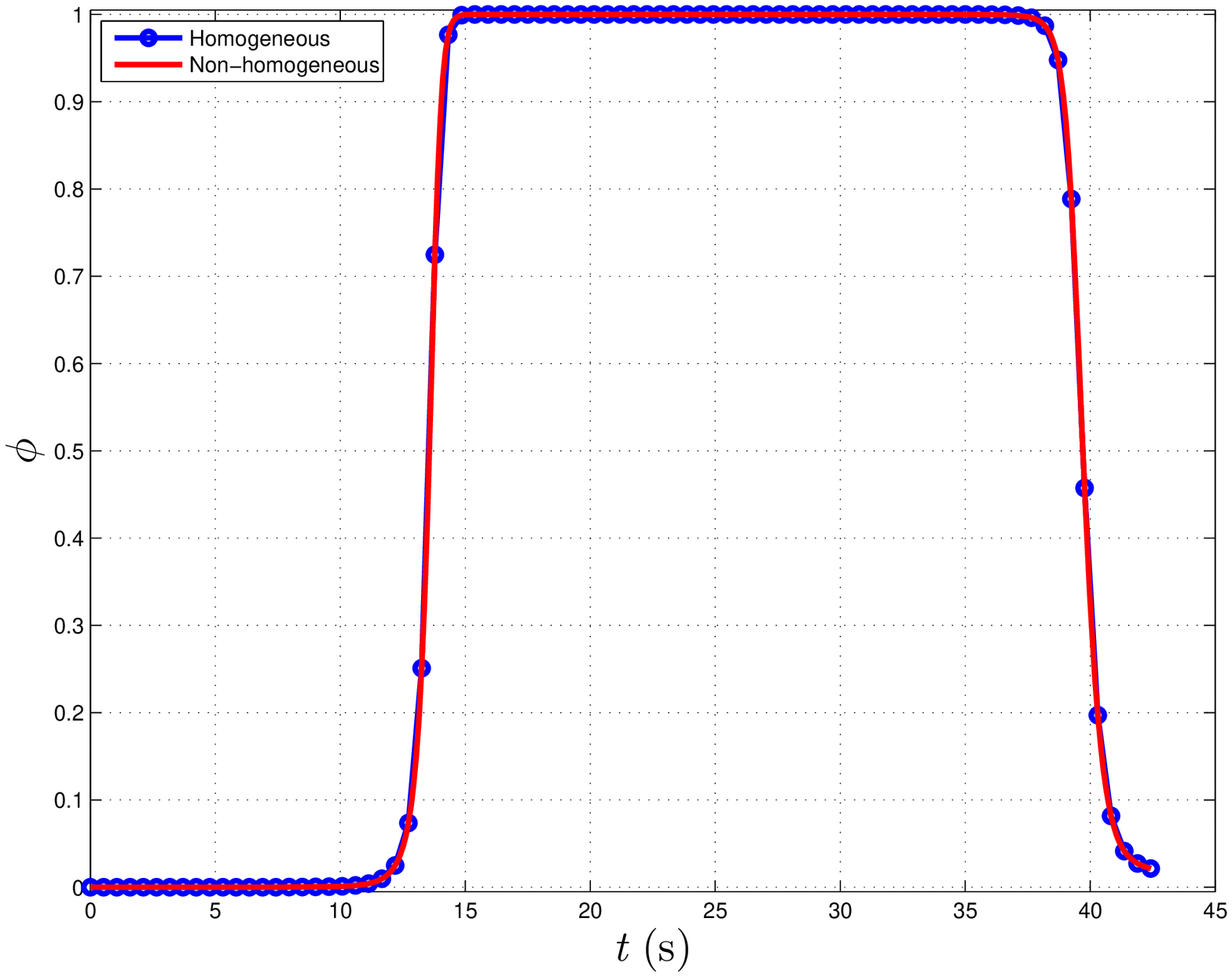}
}
\subfigure[\thetat]
{
\includegraphics[width=0.31\linewidth]{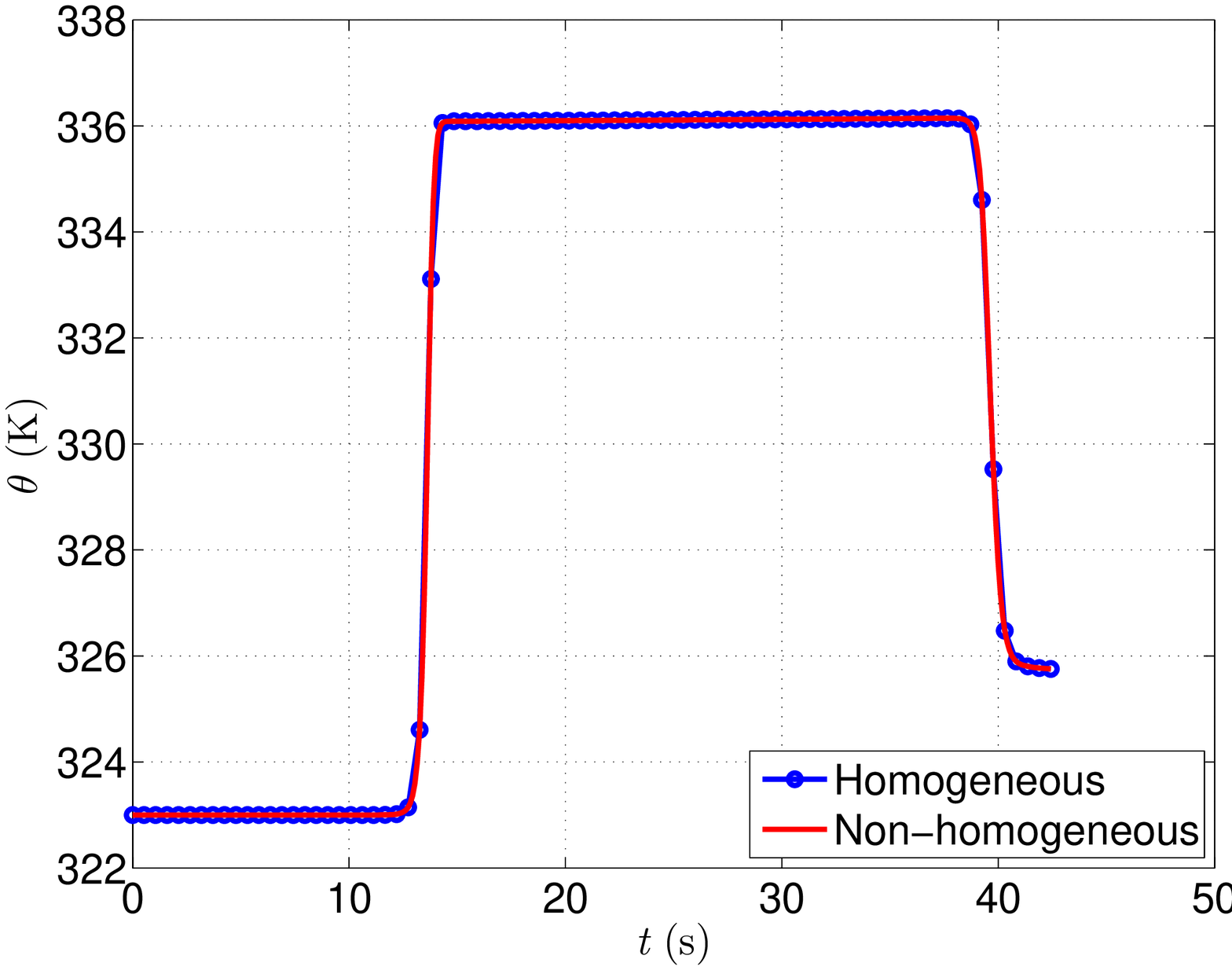}
}
\subfigure[\sxxexx]
{
\includegraphics[width=0.31\linewidth]{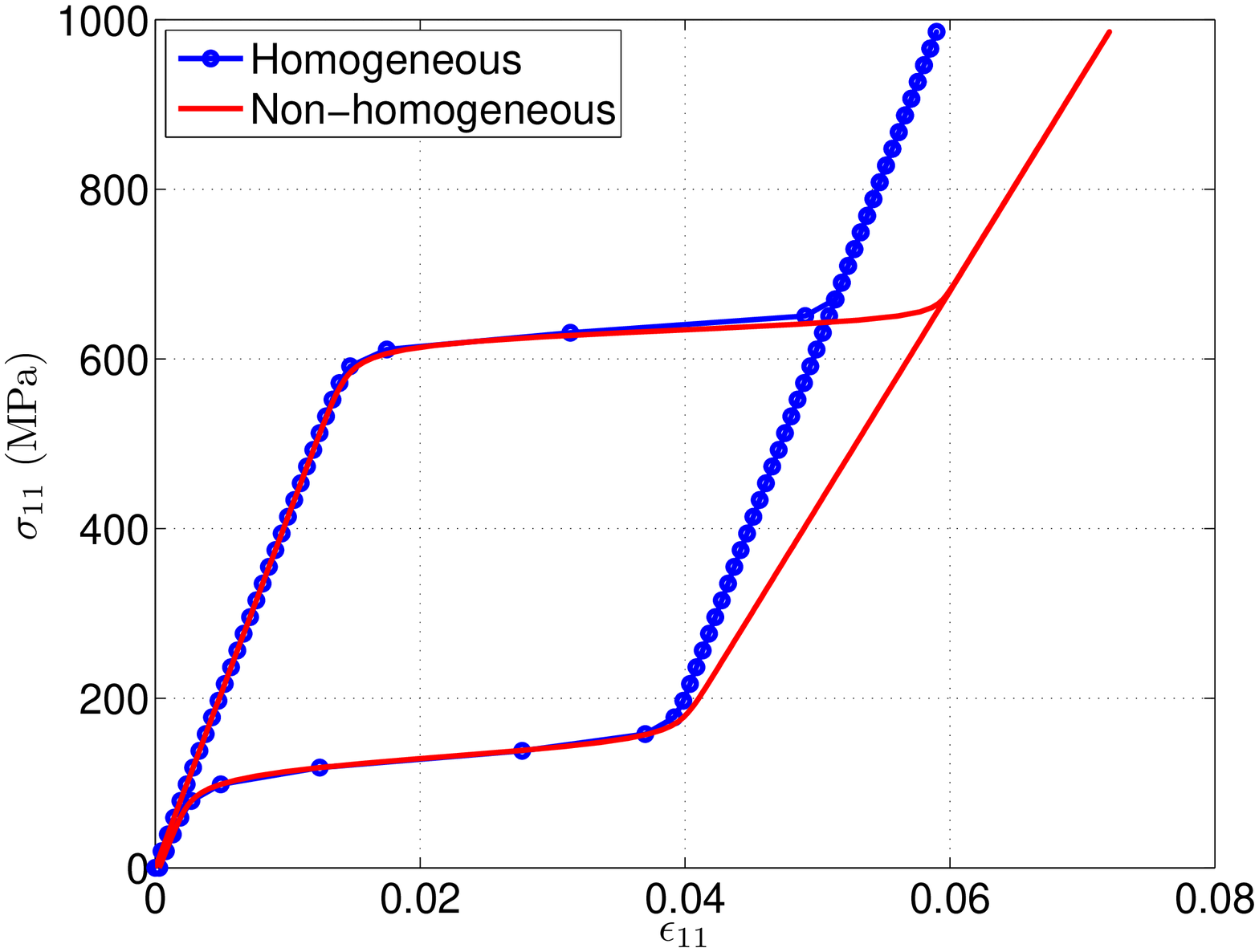}
}
\caption{(Color online) Average $\phi$, $\theta$ evolution and  \sxxexx curve for phase-dependent compliance properties.}
\label{fig:phasedependent}
\end{figure}

\begin{figure}[H]
\centering
\subfigure[]
{
\includegraphics[width=0.31\linewidth]{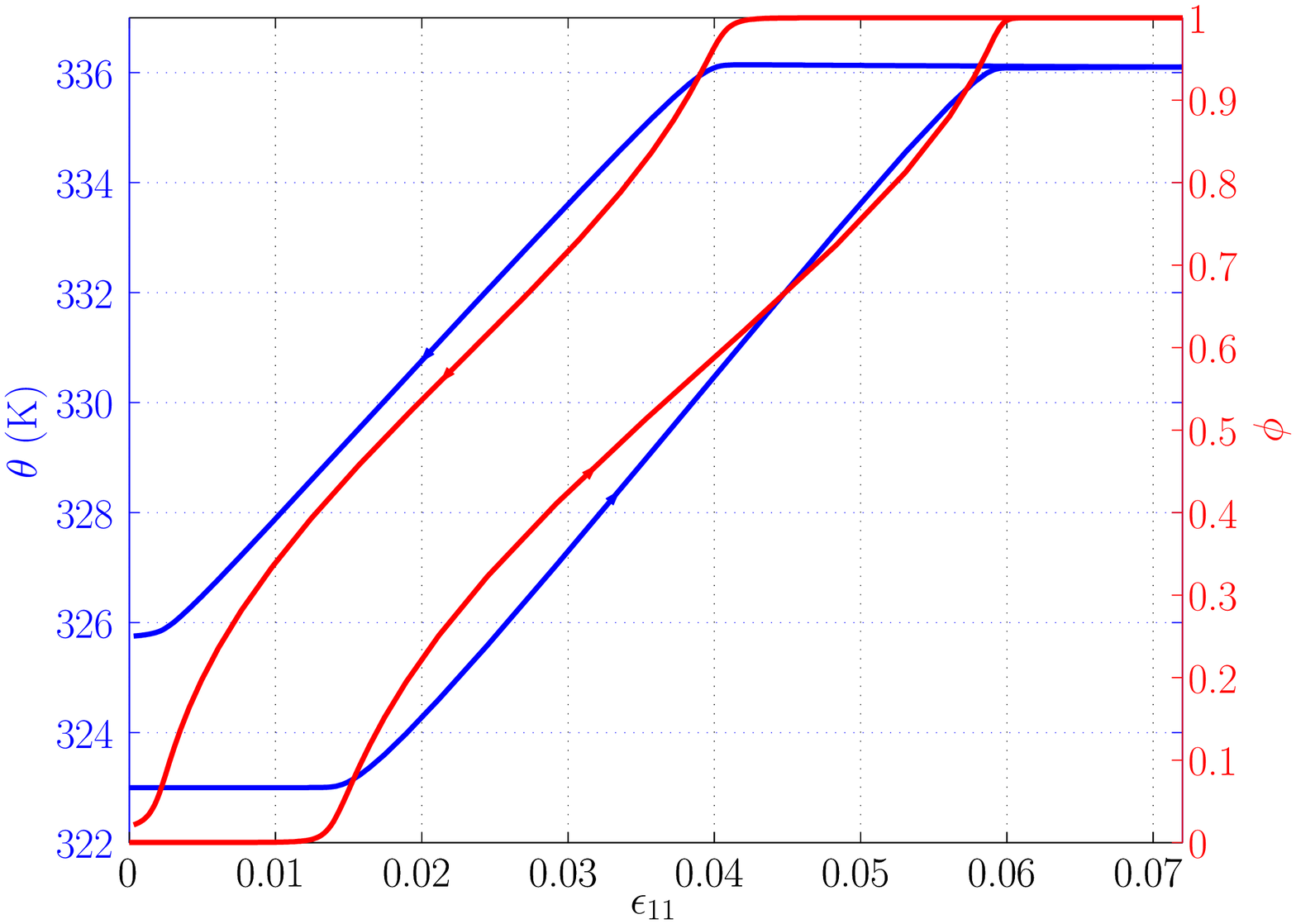}
}
\subfigure[]
{
\includegraphics[width=0.31\linewidth]{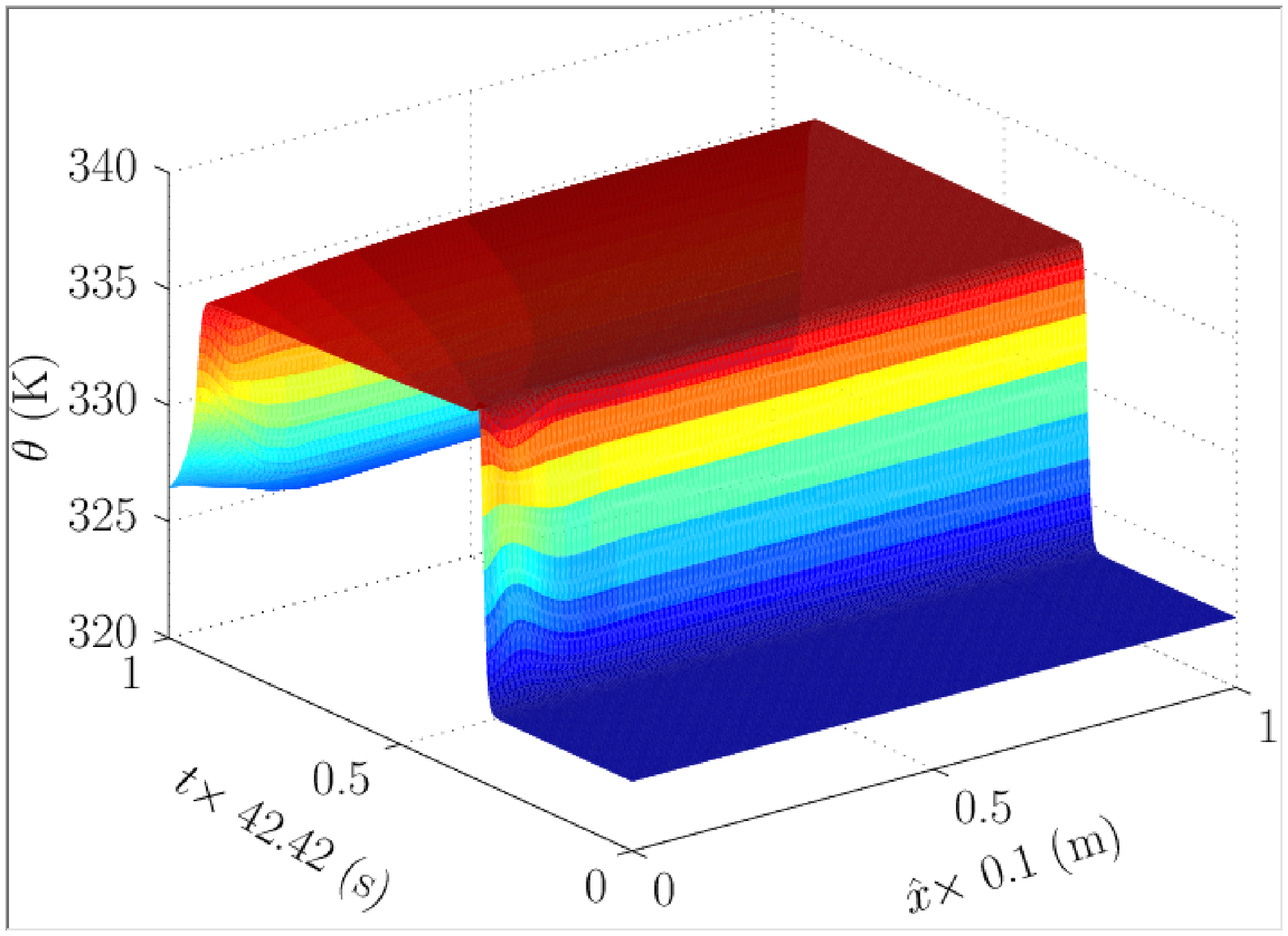}
}
\subfigure[]
{
\includegraphics[width=0.31\linewidth]{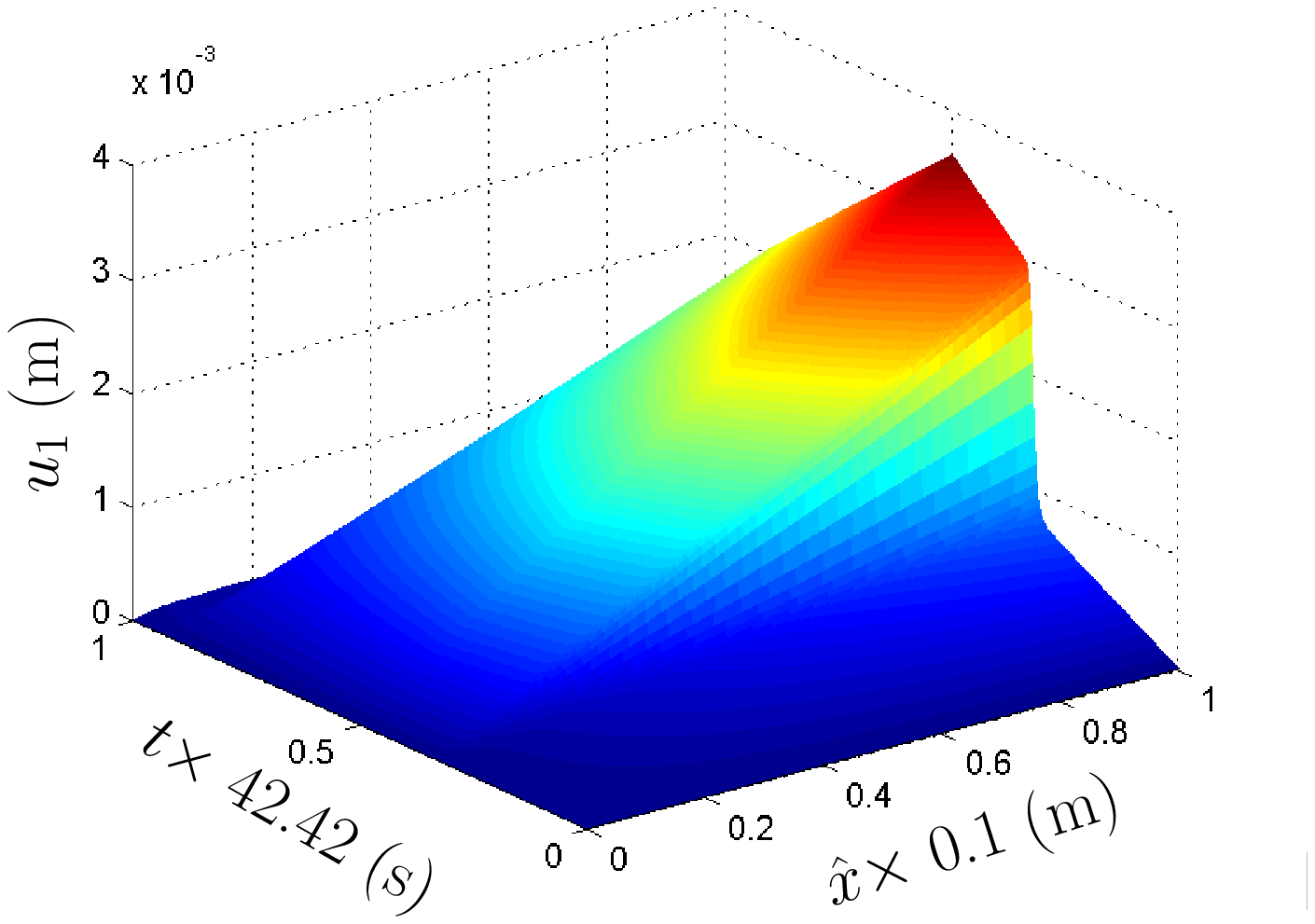}
}
\caption{(Color online) Plot of (a) $\epsilon_{11}$--$\theta$--$\phi$ loop, and the central axial line arc-length ($\hat{x}$) extrusion plot of (b) $\theta$ and (c) displacement $u_1$ evolution.}
\label{fig:phasedependentextrusion}
\end{figure}

First, a simulation has been carried out to show the effect of phase-dependent properties on the hysteretic behavior of SMAs.  Two cases are considered, first a SMA specimen with equal elastic compliances ($\lambda_M = \lambda_A$) of the two phases  and second the local phase-dependent compliance of austenite and martensite phases. The loadings are carried out with  $\dot{\sigma}_{11}$ = 46.47 MPa/s starting with initial temperature $\tilde{\theta}_{0}$ = 323 K for both the cases. The average $\phi$, $\theta$ evolution and \sxxexx curve are shown in Fig. \ref{fig:phasedependent}. The $\phi$ and $\theta$ evolve identically in both cases. However, the effect of local phase-dependent compliance is apparent on the \sxxexx curve. Figure \ref{fig:phasedependentextrusion}(a) indicates the thermal hysteresis $\theta$--$\epsilon_{11}$ and phase evolution $\phi$--$\epsilon_{11}$ loops for the local phase-dependent compliance case. The central axial line arc-length ($\hat{x}$) extrusion plot of $\theta$ and  axial displacement $u_1$ are plotted in Figs. \ref{fig:phasedependentextrusion}(b) and (c). Note that the phase-dependent compliance has been reported experimentally \cite{Otsuka,zhang2010experimental,Lagoudas}. Due to the higher compliance of martensite phase, as compared to the austenite, the area under the \sxxexx curve is high, thus causing more energy dissipation in the local phase-dependent properties. In all the subsequent simulations, we use the local phase-dependent compliance properties of the phases.

\subsection{SMA behavior at different initial temperature} \label{sec:differenttempIC}

\begin{figure}[h!]
\centering
\subfigure[]
{
\includegraphics[width=0.31\linewidth]{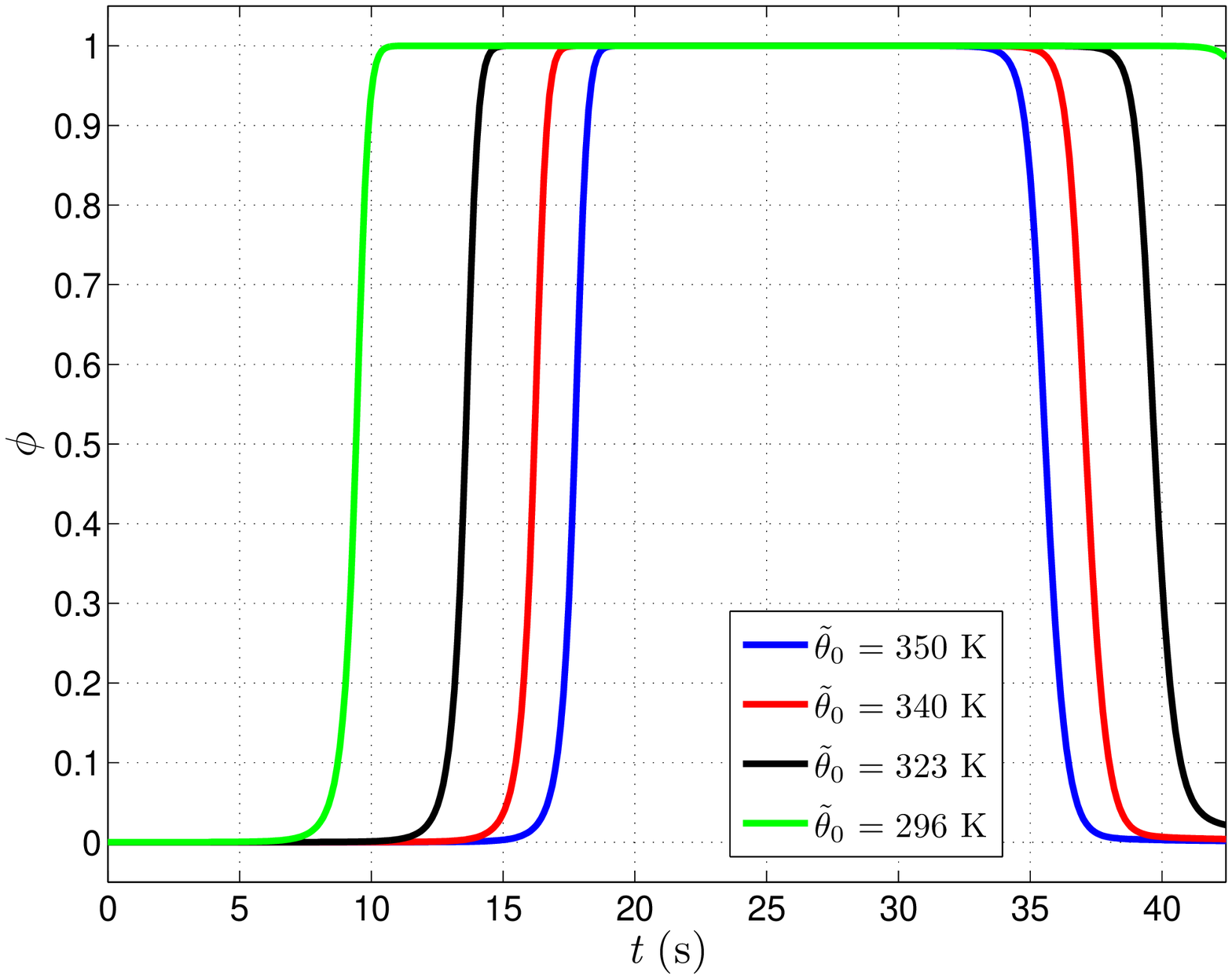}
}
\subfigure[]
{
\includegraphics[width=0.31\linewidth]{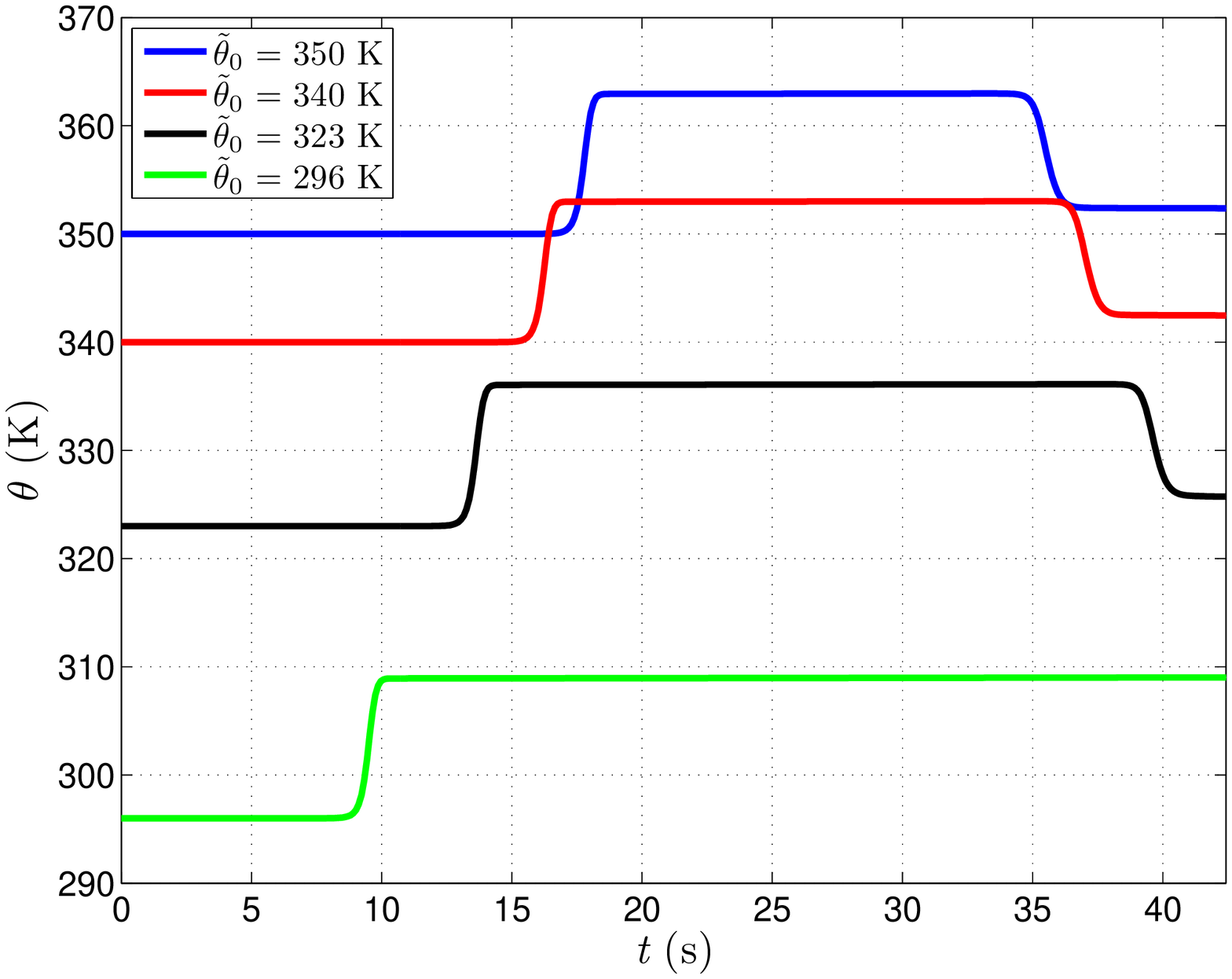}
}
\subfigure[]
{
\includegraphics[width=0.31\linewidth]{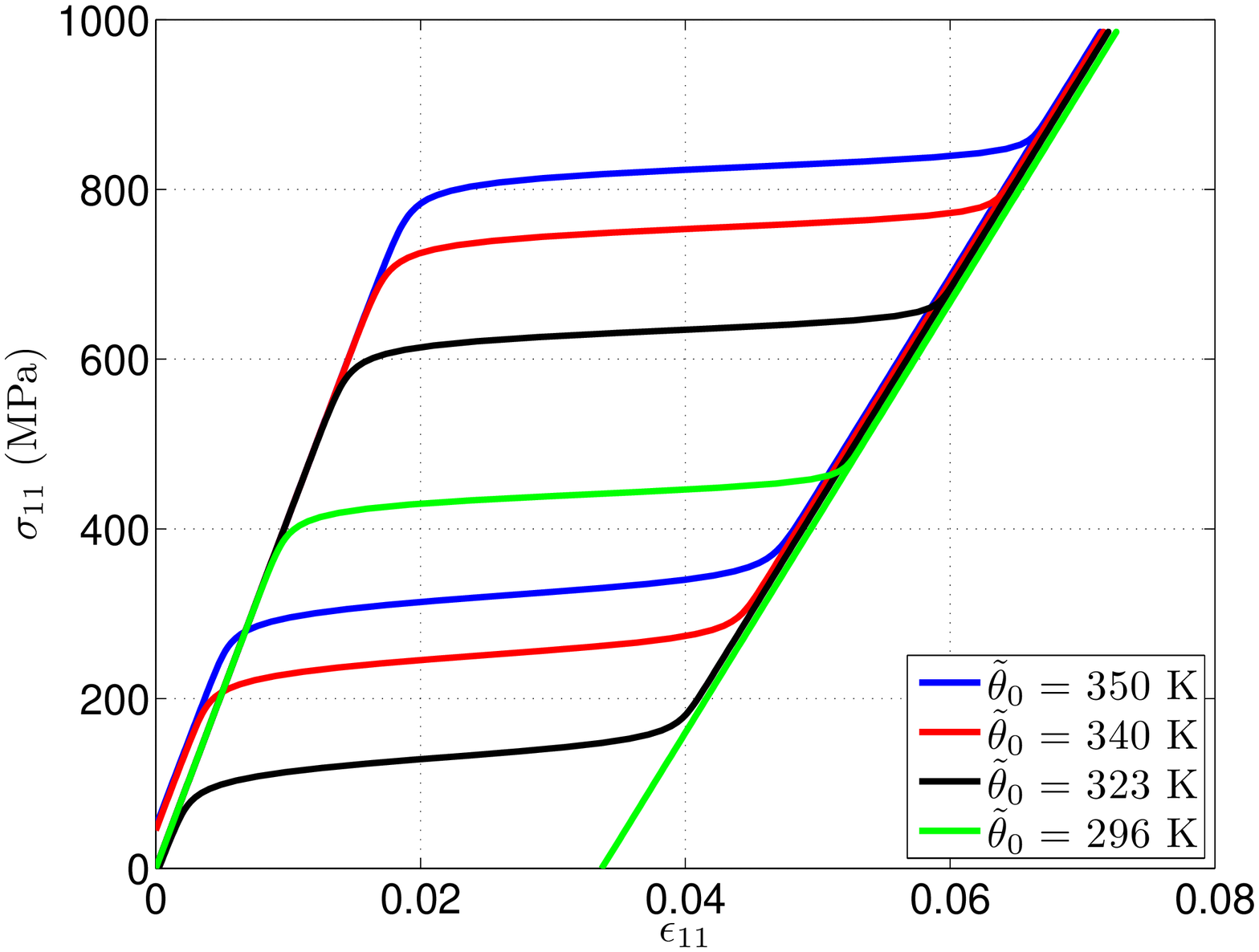}
}
\caption{(Color online) Average $\phi$, $\theta$ evolution and  \sxxexx curve starting with different $\tilde{\theta}_{0}$.}
\label{fig:differenttempIC}
\end{figure}

Next, the hysteretic response of SMA specimen is studied starting with different initial temperature $\tilde{\theta}_{0}$ ranging from 296 K to 350 K using the axial stress rate $\dot{\sigma}_{11}$ = 46.47 MPa/s. The average $\phi$, $\theta$ evolution and \sxxexx curve are shown in Fig. \ref{fig:differenttempIC}. At the lower initial temperature ($\tilde{\theta}_{0}$ = 296 K), the shape memory effect is observed with remnant strain at the end of the unloading. The pseudoelastic behavior, with fully recoverable strain, is observed at higher $\tilde{\theta}_{0}$. As $\tilde{\theta}_{0}$ increases, higher axial stress is required for the start of the phase transformation thus offsetting the \sxxexx curve towards a higher value. The area under the \sxxexx curve remains constant. These results are consistent with the results reported in the literature \cite{zhang2010experimental,Grandi2012,maraldi2012non}. 

\subsection{SMA behavior at different stress rate} \label{sec:diffstressrate}

\begin{figure}[h!]
\centering
\subfigure[]
{
\includegraphics[width=0.31\linewidth]{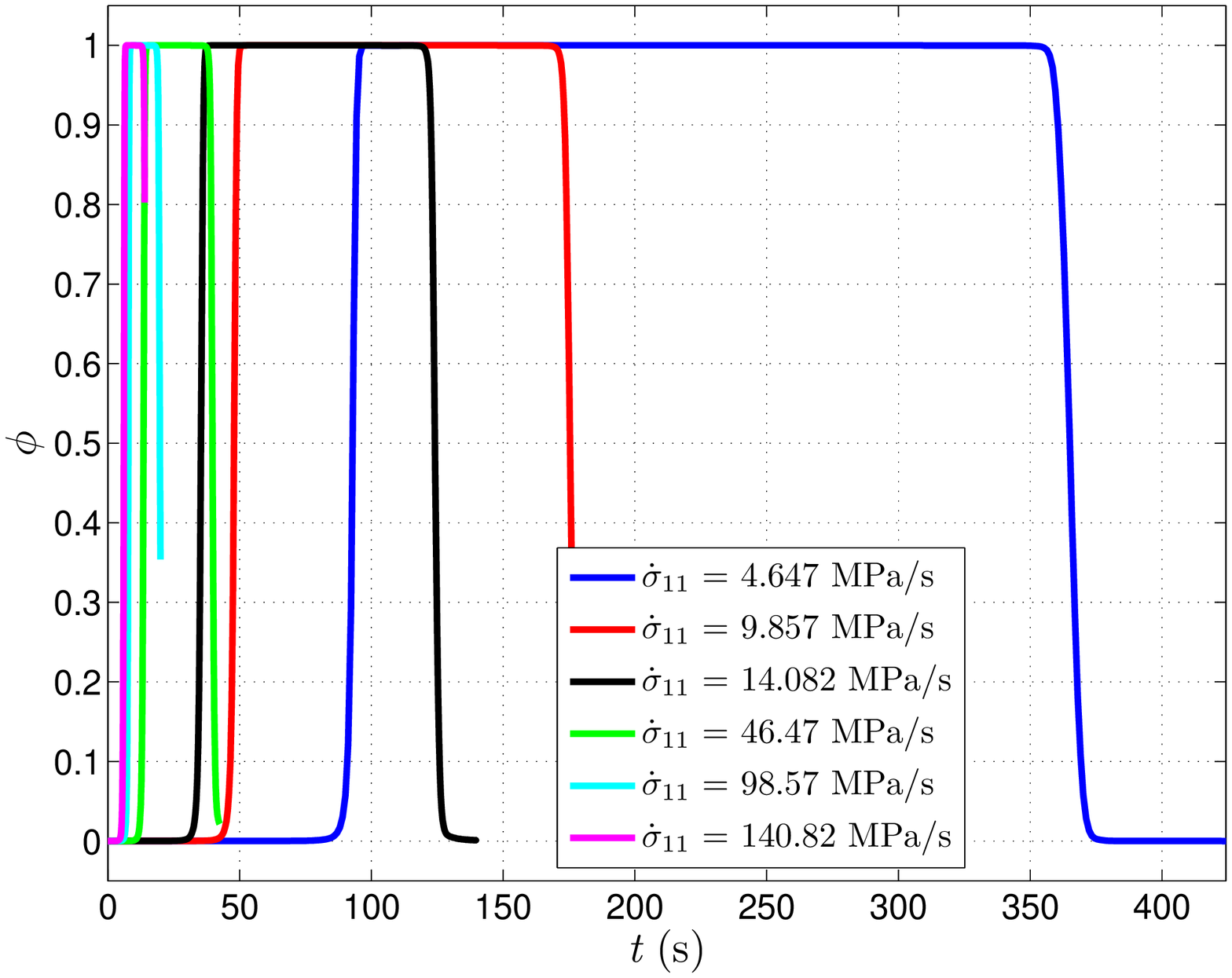}
}
\subfigure[]
{
\includegraphics[width=0.31\linewidth]{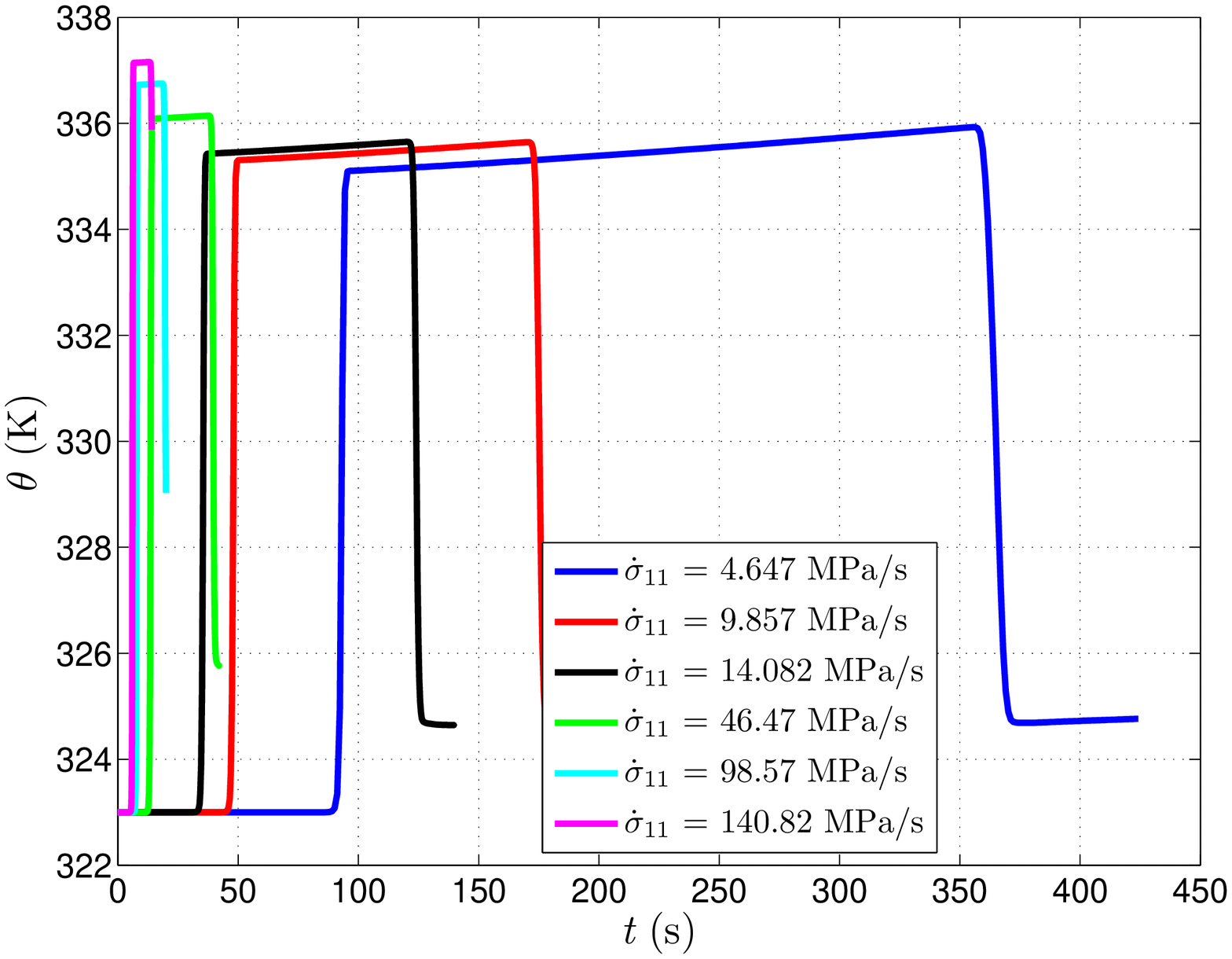}
}
\subfigure[]
{
\includegraphics[width=0.31\linewidth]{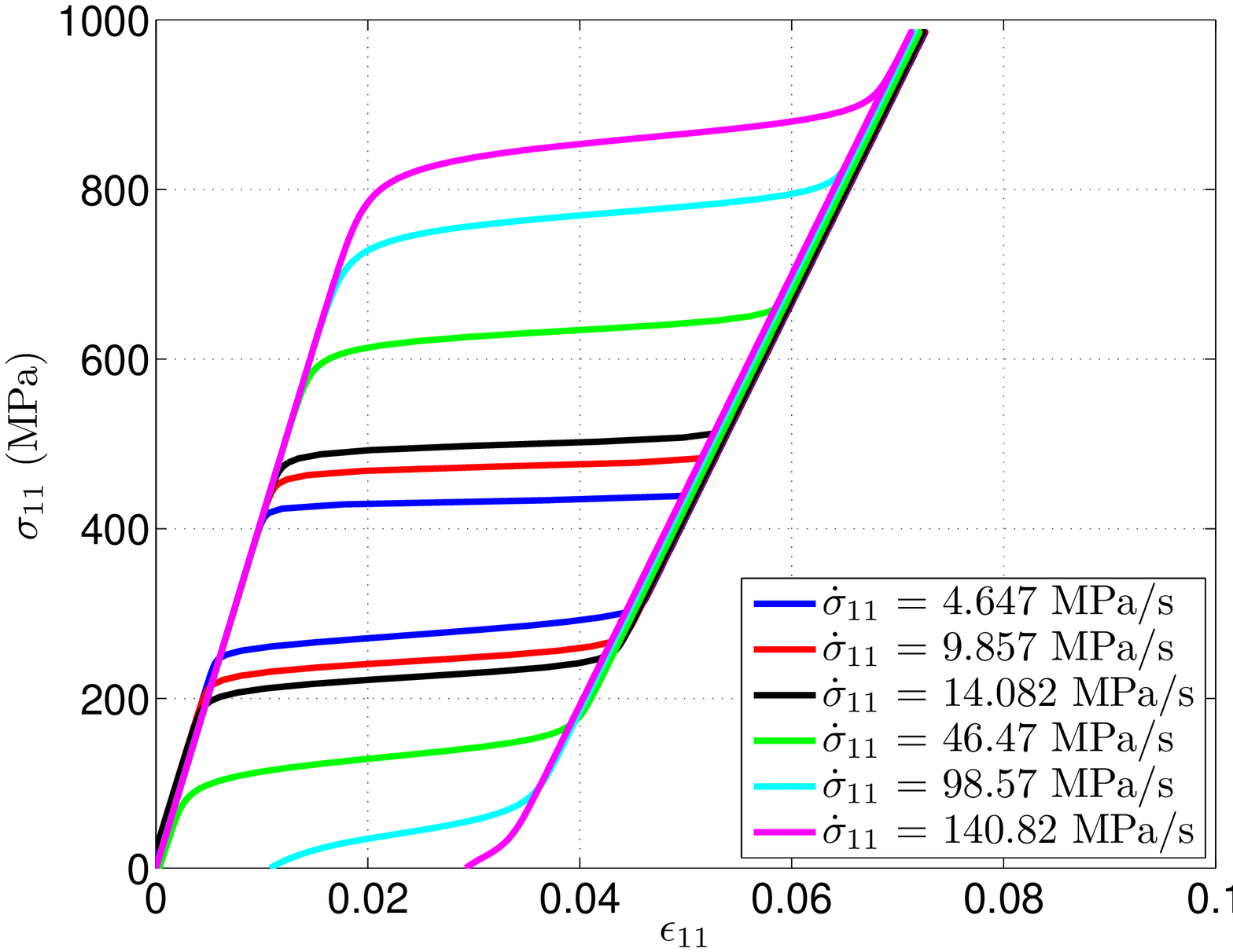}
}
\caption{(Color online) Average $\phi$, $\theta$ evolution and  \sxxexx curve for different loading rates $\dot{\sigma}_{11}$.}
\label{fig:diffstressrate}
\end{figure}

Next, the simulations have been conducted on a SMA specimen with different axial $\dot{\sigma}_{11}$ loading rates ranging between 4.647 MPa/s and 140.82 MPa/s and starting with initial temperature $\tilde{\theta}_{0}$ = 323 K. The average $\phi$, $\theta$ evolution and \sxxexx curve are shown in Fig. \ref{fig:diffstressrate}. With the increase in $\dot{\sigma}_{11}$, the phase transformation starts at higher stress value.  As the loading rate increases, the heat generated during the exothermic process \td causes the internal temperature of the specimen to increase due to insufficient time of heat transfer to the environment. During the \tdd, the heat is absorbed due to the endothermic nature of the phase transformation.  The dissipation energy (area of the hysteresis loop in the \sxxexx curve) increases with higher $\dot{\sigma}_{11}$. It is observed that with the increase in $\dot{\sigma}_{11}$, the slope of  phase transformation increases. The above behaviors have been experimentally observed \cite{zhang2010experimental}.

\section{Conclusions} \label{sec:Conclusions}

A macroscale non-isothermal 3D model has been developed to describe non-linear phase-dependent hysteretic behaviors in SMAs. The model is thermodynamically consistent with rate-dependent constitutive relationship, conservation laws of thermal and mechanical physics in conjunction with a kinetic phase evolution equation. In its description of the phase transformations, the model is based on the stress tensor and utilizes the scalar order parameter. 

The representative numerical simulations for the stress controlled loadings illustrate that the phase-dependent compliance properties improve the pseudoelastic non-linear hysteretic description. The tensile tests on SMA specimen with different initial temperatures, as well as loading stress rates, elucidate the model's ability to capture efficiently the thermo-mechanical behavior and phase kinetics. The model successfully reproduces experimentally observed SMA behaviors reported in the literature. 

\section*{Acknowledgement}
RD and RM have been supported by NSERC and CRC program, Canada.

\begingroup
\setstretch{0.8}
\bibliographystyle{unsrt}
\bibliography{NonisothermalarXiv}

\begin{thebibliography}{10}

\bibitem{Otsuka}
K.~Otsuka and C.~Wayman.
\newblock {\em {Shape memory materials}}.
\newblock Cambridge University Press, 1998.

\bibitem{Kohl2004}
M.~Kohl.
\newblock {\em Shape memory microactuators}.
\newblock Springer Verlag, 2004.

\bibitem{Lagoudas}
D.~Lagoudas.
\newblock {\em {Shape Memory Alloys: Modeling and Engineering Applications}}.
\newblock Springer, London, 2008.

\bibitem{Miyazaki2009}
S.~Miyazaki, Y.Q. Fu, and W.M. Huang.
\newblock {\em Thin film shape memory alloys: fundamentals and device
  applications}.
\newblock Cambridge University Press, 2009.

\bibitem{Ozbulut2011}
O.~Ozbulut, S.~Hurlebaus, and R.~DesRoches.
\newblock Seismic response control using shape memory alloys: A review.
\newblock {\em Journal of Intelligent Material Systems and Structures},
  22(14):1531--1549, 2011.

\bibitem{Elahinia2011}
M.~Elahinia, M.~Hashemi, M.~Tabesh, and S.~Bhaduri.
\newblock {Manufacturing and processing of NiTi implants: a review}.
\newblock {\em Progress in Materials Science}, 2011.

\bibitem{Khandelwal2009}
A.~Khandelwal and V.~Buravalla.
\newblock {Models for Shape Memory Alloy Behavior: An overview of modelling
  approaches}.
\newblock {\em International Journal of Structural Changes in Solids -
  Mechanics and Applications}, 1(1):111--148, 2009.

\bibitem{Lagoudas2006}
D.~Lagoudas, L.~Brinson, and E.~Patoor.
\newblock {Shape memory alloys, Part II: Modeling of polycrystals}.
\newblock {\em Mech. Mater.}, 38(5-6):430--462, 2006.

\bibitem{Birman1997}
V.~Birman.
\newblock {Review of mechanics of shape memory alloy structures}.
\newblock {\em Applied Mechanics Reviews}, 50(11):629--645, 1997.

\bibitem{Smith2005}
R.C. Smith.
\newblock {\em Smart material systems: model development}, volume~32.
\newblock Society for Industrial Mathematics, 2005.

\bibitem{Paiva2006}
A.~Paiva and M.A. Savi.
\newblock An overview of constitutive models for shape memory alloys.
\newblock {\em Mathematical Problems in Engineering}, 2006:1--30, 2006.

\bibitem{mamivand2013review}
M.~Mamivand, M.~Zaeem, and H.~El~Kadiri.
\newblock A review on phase field modeling of martensitic phase transformation.
\newblock {\em Computational Materials Science}, 77:304--311, 2013.

\bibitem{Falk1980}
F.~Falk.
\newblock {Model free energy, mechanics, and thermodynamics of shape memory
  alloys}.
\newblock {\em Acta Metallurgica}, 28(12):1773--1780, 1980.

\bibitem{Khachaturian1983}
A.~Khachaturian.
\newblock {\em Theory of structural transformations in solids}.
\newblock John Wiley and Sons, New York, NY, 1983.

\bibitem{Melnik1999}
R.~Melnik, A.~Roberts, and K.~Thomas.
\newblock Modelling dynamics of shape-memory-alloys via computer algebra.
\newblock {\em Proc. of SPIE Mathematics and Control in Smart Structures},
  3667:290--301, 1999.

\bibitem{Melnik2000}
R.~Melnik, A.~Roberts, and K.~A. Thomas.
\newblock {Computing dynamics of Copper-based SMA via center manifold reduction
  models}.
\newblock {\em Computational Material Science}, 18:255--268, 2002.

\bibitem{Artemev2001}
A.~Artemev, Y.~Jin, and A.G. Khachaturyan.
\newblock Three-dimensional phase field model of proper martensitic
  transformation.
\newblock {\em Acta Materialia}, 49(7):1165 -- 1177, 2001.

\bibitem{Chen2002}
L.~Chen.
\newblock {Phase Field Models for Microstructure Evolution}.
\newblock {\em Annual Review of Materials Research}, 32:113--140, 2002.

\bibitem{Levitas2002a}
V.~Levitas and D.~Preston.
\newblock {Three-dimensional Landau theory for multivariant stress-induced
  martensitic phase transformations. I. austenite $\leftrightarrow$
  martensite}.
\newblock {\em Physics Review B}, 66(134206):1--9, 2002.

\bibitem{Levitas2002b}
V.~Levitas and D.~Preston.
\newblock {Three-dimensional Landau theory for multivariant stress-induced
  martensitic phase transformations. II. Multivariant phase transformations and
  stress space analysis}.
\newblock {\em Physics Review B}, 66(134206):1--15, 2002.

\bibitem{Levitas2003}
V.~Levitas, D.~Preston, and D.~Lee.
\newblock {Three-dimensional Landau theory for multivariant stress-induced
  martensitic phase transformations. III. Alternative potentials, critical
  nuclei, kink solutions, and dislocation theory}.
\newblock {\em Physics Review B}, 68(134201):1--24, 2003.

\bibitem{Ahluwalia2006}
R.~Ahluwalia, T.~Lookman, and A.~Saxena.
\newblock {Dynamic Strain Loading of Cubic to Tetragonal Martensites}.
\newblock {\em {Acta Mater.}}, 54:2109--2120, 2006.

\bibitem{Mahapatra2006}
D.~Mahapatra and R.~Melnik.
\newblock \textrm{Finite Element Analysis of Phase Transformation Dynamics in
  Shape Memory Alloys with a Consistent Landau-Ginzburg Free Energy Model }.
\newblock {\em Mechanics of Advanced Materials and Structures}, 13(6):443 --
  455, 2006.

\bibitem{Mahapatra2007a}
D.~Mahapatra and R.~Melnik.
\newblock {Finite element approach to modelling evolution of $3$D shape memory
  materials}.
\newblock {\em Mathematics and Computers in Simulation}, 76:141--148, 2007.

\bibitem{wang2007finite}
L.X. Wang and R.V.N. Melnik.
\newblock Finite volume analysis of nonlinear thermo-mechanical dynamics of
  shape memory alloys.
\newblock {\em Heat and mass transfer}, 43(6):535--546, 2007.

\bibitem{Bouville2008}
M.~Bouville and R.~Ahluwalia.
\newblock {Microstructure and Mechanical Properties of Constrained Shape Memory
  Alloy Nanograins and Nanowires}.
\newblock {\em Acta Mater.}, 56(14):3558--3567, 2008.

\bibitem{Bouville2009}
M.~Bouville and R.~Ahluwalia.
\newblock {Phase field simulations of coupled phase transformations in
  ferroelastic-ferroelastic nanocomposites}.
\newblock {\em Phys. Rev. B}, 79(9):094110, 2009.

\bibitem{Grandi2012}
D.~Grandi, M.~Maraldi, and L.~Molari.
\newblock A macroscale phase-field model for shape memory alloys with
  non-isothermal effects: Influence of strain rate and environmental conditions
  on the mechanical response.
\newblock {\em Acta Materialia}, 60(1):179--191, 2012.

\bibitem{Dhote2012}
R.~Dhote, R.~Melnik, and J.~Zu.
\newblock Dynamic thermo-mechanical coupling and size effects in finite shape
  memory alloy nanostructures.
\newblock {\em Computational Materials Science}, 63:105--117, 2012.

\bibitem{Dhote2013}
R.~Dhote, M.~Fabrizio, R.~Melnik, and J.~Zu.
\newblock Hysteresis phenomena in shape memory alloys by non-isothermal
  ginzburg-landau models.
\newblock {\em Communications in Nonlinear Science and Numerical Simulation},
  18:2549--2561, 2013.

\bibitem{wang1997three}
Y.~Wang and AG~Khachaturyan.
\newblock Three-dimensional field model and computer modeling of martensitic
  transformations.
\newblock {\em Acta Mater.}, 45(2):759--773, 1997.

\bibitem{curnoe2000twin}
S.~Curnoe and A.~Jacobs.
\newblock Twin wall of proper cubic-tetragonal ferroelastics.
\newblock {\em Physical Review B}, 62(18):11925--11928, 2000.

\bibitem{Lookman2003}
T.~Lookman, S.~R. Shenoy, K.~Rasmussen, A.~Saxena, and A.~R. Bishop.
\newblock {Ferroelastic dynamics and strain compatibility}.
\newblock {\em Phys. Rev. B Condens. Matter. Mater. Phys.}, 67(2):24114, 2003.

\bibitem{Berti2010}
V.~Berti, M.~Fabrizio, and D.~Grandi.
\newblock {Phase transitions in shape memory alloys: A non-isothermal Ginzburg
  Landau model}.
\newblock {\em Physica D: Nonlinear Phenomena}, 239(1-2):95--102, 2010.

\bibitem{Berti2010a}
V.~Berti, M.~Fabrizio, and D.~Grandi.
\newblock {Hysteresis and phase transitions for one-dimensional and
  three-dimensional models in shape memory alloys}.
\newblock {\em Journal of Mathematical Physics}, 51(6):062901, 2010.

\bibitem{grandi2012macroscale}
D.~Grandi, M.~Maraldi, and L.~Molari.
\newblock A macroscale phase-field model for shape memory alloys with
  non-isothermal effects: Influence of strain rate and environmental conditions
  on the mechanical response.
\newblock {\em Acta Mater.}, 60(1):179--191, 2012.

\bibitem{maraldi2012non}
M.~Maraldi, L.~Molari, and D.~Grandi.
\newblock A non-isothermal phase-field model for shape memory alloys: Numerical
  simulations of superelasticity and shape memory effect under
  stress-controlled conditions.
\newblock {\em J. Intel. Mat. Syst. Str.}, 23(10):1083--1092, 2012.

\bibitem{zhang2010experimental}
X.~Zhang, P.~Feng, Y.~He, T.~Yu, and Q.~Sun.
\newblock Experimental study on rate dependence of macroscopic domain and
  stress hysteresis in niti shape memory alloy strips.
\newblock {\em International Journal of Mechanical Sciences},
  52(12):1660--1670, 2010.

\bibitem{Comsol}
{Comsol Multiphysics Software, Version 4.2}.
\newblock (\url{http://www.comsol.com}).

\end{thebibliography}
\endgroup

\end{document}